\documentclass[12pt,letterpaper]{article}
\usepackage[includeheadfoot, 
            marginratio={1:1,2:3}, 
            width=412pt, 
            height=689pt,]{geometry}

\usepackage{amsmath}
\usepackage{amsfonts}
\usepackage{amssymb}

\usepackage{graphicx}
\usepackage{subfigure}

\allowdisplaybreaks[2]
\numberwithin{equation}{section}

\newcommand{\eq}[1]{\begin{equation}
                     \begin{split} #1 \end{split}
                     \end{equation}}

\def\ov{\overline}

\begin{document}


\vspace*{-1.5cm}
\begin{flushright}
  {\small
  MPP-2008-60
  }
\end{flushright}

\vspace{2cm}
\begin{center}
  {\LARGE
  String GUT Scenarios with Stabilised Moduli
  }
\end{center}

\vspace{0.75cm}
\begin{center}
  Ralph Blumenhagen, Sebastian Moster, Erik Plauschinn \\
\end{center}

\vspace{0.1cm}
\begin{center}
  Max-Planck-Institut f\"ur Physik \\ F\"ohringer Ring 6 \\
  80805 M\"unchen\\ Germany 
\end{center}

\vspace{-0.1cm}
\begin{center}
  \tt{
  blumenha, moster, plausch @mppmu.mpg.de \\
  }
\end{center}

\vspace{1.5cm}
\begin{abstract}
\noindent
Taking into account the recently proposed poly-instanton corrections to the
superpotential and
combining the race-track with a KKLT respectively LARGE Volume Scenario
in an intricate manner, we show that we gain 
exponential control
over the parameters in an effective superpotential. This allows
us to dynamically stabilise moduli such that 
a conventional MSSM scenario with the string scale 
lowered to the GUT scale is realised.
Depending on the cycles wrapped by the MSSM branes, two different
scenarios for the hierarchy of soft masses 
arise. The first one is a supergravity mediated model with $M_{3/2}\simeq 1\,$TeV while the second one features mixed anomaly-supergravity mediation
with $M_{3/2}\simeq 10^{10}\,$GeV and split supersymmetry.
We also comment on dynamically lowering the scales such that
the tree-level cosmological constant is of the order 
$\Lambda=(10^{-3} {\rm eV})^4$.
\end{abstract}

\thispagestyle{empty}
\clearpage
\tableofcontents


\section{Introduction}

The origin of the non-vanishing but extremely small positive cosmological 
constant $\Lambda\simeq (10^{-3}\,{\rm eV})^4$, as 
deduced from precision measurements of the cosmic microwave background, 
is still elusive. Similarly, the origin and stability of the weak scale
are still not clearly understood, although we hope to find 
evidence for an answer at the LHC. 
Various possibilities to explain 
the weak scale as well as the gauge hierarchy problem have
been proposed, the most prominent surely being supersymmetry
which is supported by gauge coupling
unification at the scale $M_X=1.2\cdot 10^{16}\,{\rm GeV}$ within the MSSM.
String theory motivates also more exotic ideas descending from the
presence of extra space-dimensions such as Large Extra Dimensions
\cite{ArkaniHamed:1998rs,Antoniadis:1998ig} (see also \cite{Antoniadis:1990ew})
or the Randall-Sundrum \cite{Randall:1999ee} scenario.

For string theory to make contact with low energy phenomenology, the 
problem of moduli stabilisation has to be addressed. In fact, during
the last years considerable progress has been made. In particular,
KKLT \cite{Kachru:2003aw} proposed a scenario which led to meta-stable de Sitter vacua with all moduli stabilised. In this scenario, the complex structure moduli are 
fixed at tree-level  by fluxes while the K\"ahler moduli are stabilised via non-perturbative contributions to the superpotential
\begin{equation}
  \label{kklt}
  W=W_0+A\, e^{-\alpha\, T} \;.
\end{equation}
As a generalisation of this, including also
next to leading order corrections to the K\"ahler potential, 
the volume of the compactification manifold can be
stabilised at exponentially
large values \cite{Balasubramanian:2005zx}. These large volume minima are quite generic \cite{Cicoli:2008va}
and have been called  LARGE Volume Scenarios (LVS).
Their phenomenological features were studied
in very much detail for the string scale in the intermediate regime
$M_{\rm s}\simeq 10^{11}\,{\rm GeV}$ \cite{Conlon:2005ki,Conlon:2007xv}
leading to intermediate scales for the neutrino
and axion sector of the MSSM.

In this paper, we further extend the KKLT respectively LARGE Volume
Scenario by including the recently proposed poly-instanton
corrections \cite{Blumenhagen:2008ji} to the superpotential. 
Let us emphasise that such corrections
are of genuine stringy origin and 
are not expected to be understood from pure field theory.  
With these contributions it is quite simple to combine
a race-track scenario with the LVS
allowing for exponential control over effective
parameters $W^{\rm eff}_0$ and $A^{\rm eff}$ in a superpotential
such as \eqref{kklt}.
In particular, the stabilised race-track modulus controls the size of $W^{\rm eff}_0$ and $A^{\rm eff}$ allowing for exponential small values without fine-tuning.\footnote{Moduli stabilisation via
race-track type superpotentials in the context of M-theory
has been discussed in \cite{Acharya:2007rc}.}

One particular scenario we are aiming for 
is a conventional supersymmetric Grand Unified Theory (GUT) with
MSSM  gauge coupling unification at $M_X$.
Computing the resulting soft masses for  MSSM branes supported on 
small cycles of the Calabi-Yau manifold suggests two different
setups.
One is very similar to the LVS featuring a gravitino
mass around the TeV scale with a supergravity mediated 
$\ln(M_{\rm Pl}/M_{3/2})$ suppression of the soft-terms.
The second scenario utilises the observation that the
race-track modulus is fixed in an almost supersymmetric minimum.
The soft masses on $D7$-branes wrapping a four-cycle corresponding to this modulus are automatically
hierarchically suppressed. 
In particular, the gravitino and soft scalar masses are obtained in an intermediate regime
while the gaugino masses are found at the weak scale, dominantly generated
by anomaly mediation.

The outline of this paper is as follows. In section \ref{sec_mod_stab} we define and analyse our scenario which is a combination
of a supersymmetric flux minimum and a race-track superpotential
subject to string instanton corrections. 
In section \ref{sec_gut}, 
we present supersymmetric GUT scenarios
with $M_{\rm s}=M_X$ where realistic gauge coupling unification arises dynamically,
and in section \ref{sec_cc} we comment on the cosmological
constant problem. 
Finally, in section \ref{sec_concl} we summarise our findings.


\section{Moduli Stabilisation}
\label{sec_mod_stab}

We consider type IIB orientifolds of Calabi-Yau manifolds 
with $O7$- and $O3$-planes which are the currently best understood 
framework for moduli stabilisation.   
In order to cancel the $C_8$ and induced $C_4$ tadpoles, 
we introduce $D7$- 
and $D3$-branes as well as the combination
$G_3=F_3+S\,  H_3$ of R-R and NS-NS fluxes.


\bigskip
The $G_3$-flux gives rise to a tree-level superpotential of the form \cite{Gukov:1999ya} 
\begin{equation}
  \label{w_GVW}
  W_{\rm{flux}} = \int_{\mathcal{X}} G_3\wedge \Omega_3 \;,
\end{equation}
which in general allows to stabilise all complex structure 
moduli $U_i$ together with the axio-dilaton $S$ by the 
supersymmetry conditions $D_{U_i}W_{\rm{flux}}=D_S W_{\rm{flux}}=0$.
In contrast to KKLT and LVS,
we require in addition that 
\eq{
  \label{flux_zero}
  W_{\rm flux}\,\bigr\rvert_{\rm min.}=0 \;.
}
This is an over-constrained system, but it was shown 
in \cite{DeWolfe:2004ns} that such solutions
are not highly suppressed. In fact, it was provided evidence that 
the number of minima \eqref{flux_zero} compared to all flux minima scales as $\#(W_{\rm flux}\rvert_{\rm min.}=0)/\#({\rm tot})\sim 1/L^{\mathcal{D}/2}$ with 
$L$ the upper limit on the flux quanta and $\mathcal{D}$ an integer.

Finally, we note that $D_{U_i}W$ and $D_S W$ appear at order $\mathcal{V}^{-2}$ in the scalar potential while the K\"ahler moduli, as we will see below, appear at orders $\mathcal{V}^{-1}$ and $\mathcal{V}^{-3}$. In the limit of large $\mathcal{V}$ we are interested in, we can therefore stabilise $U_i$ and $S$ independently of the K\"ahler moduli. 


\bigskip
Let us now study in more detail the K\"ahler
moduli $T_a=\tau_a+i\,\rho_a$ with $\tau_a$  the four-cycle volumes of the compact space
and $\rho_a$ the axions originating from $C_4$. We compute the F-term potential 
using the standard K\"ahler potential including
$\alpha'$-corrections \cite{Becker:2002nn} 
\begin{equation}
  \label{kahl-pot}
  \mathcal{K} = -2\ln \Bigl( \mathcal{V} + {\textstyle \frac{{\hat\xi}}{2} }
  \Bigr) 
  - \ln \Bigl( S+\overline{S} \Bigr)
  + \mathcal{K}_{\rm CS}
\end{equation}
where $\hat\xi=\xi/g_s^{3/2}$ and $g_s$ is the string coupling.
The resulting inverse K\"ahler metric for the K\"ahler moduli reads
\begin{equation}
  G^{a \ov b}
  =-2\,\Bigl( \mathcal{V} + {\textstyle \frac{{\hat\xi}}{2} } \Bigr)  
  \biggl( \frac{\partial^2 \mathcal{V}}
    {\partial\tau_a\,\partial\tau_b} \biggr)^{-1} 
  +\tau_a\, \tau_b\, 
    \frac{4\,\mathcal{V}-\hat\xi}{\mathcal{V}-\hat\xi }\; ,
\end{equation}
where the inverse denotes the usual matrix inverse 
and we choose the volume $\mathcal{V}$ of the internal space 
to be of {\em swiss-cheese} form with three K\"ahler moduli 
\begin{equation}
  \label{vol1}
  \mathcal{V} =  \bigl( \eta_{\rm b} \tau_{\rm b}\bigr)^{3/2} 
  - \bigl(\eta_1 \tau_1 \bigr)^{3/2}
  - \bigl(\eta_2 \tau_2 \bigr)^{3/2} \;
\end{equation}
Here, $\tau_{\rm b}$ controls the overall volume $\mathcal{V}$ and $\tau_{(1,2)}$ are small holes in this geometry.
The constants $\eta_{\rm b},\eta_1,\eta_2$ are determined by a specific choice of a compactification manifold.


Furthermore, we assume gaugino condensation 
on two stacks of $D7$-branes
wrapping the four-cycle $\Gamma_1$ corresponding to the K\"ahler modulus $T_1$.\footnote{On the type I side, such a setup can be realised for instance by discrete Wilson lines
as it has been shown in \cite{Camara:2007dy}.} 
This leads to a race-track superpotential containing
exponentials of the two gauge kinetic functions.
In further developing the D-brane instanton calculus
pioneered in \cite{Billo:2002hm,Blumenhagen:2006xt,Ibanez:2006da,Florea:2006si,Akerblom:2006hx,Bianchi:2007fx,Cvetic:2007ku,Argurio:2007vqa,Bianchi:2007wy,Ibanez:2007rs}, 
it has been argued in \cite{Akerblom:2007uc,Blumenhagen:2008ji}
that also the gauge kinetic function receives 
non-perturbative corrections
from euclidean $D3$-brane instantons. 
For such an instanton to contribute, it
must have a  zero-mode structure specified by 
$h_{2,0}(\Gamma_{E3})=1$ and $h_{1,0}(\Gamma_{E3})=0$
where $\Gamma_{E3}$ denotes the cycle wrapped by the instanton. 
Let us assume that such corrections
can indeed arise from an instanton wrapping the four-cycle $\Gamma_2$ 
with K\"ahler modulus $T_2$. 
Note that, because of its zero-mode structure, this instanton will not contribute as a single instanton and 
so the superpotential at leading order reads (see also \cite{Grimm:2007xm})
\begin{align}
  \label{instanton}
  W_{\rm np}
    &= \hspace{6.5pt}
      \mathcal{A}\, 
      e^{-a\,\left( T_1 + \mathcal{C}_1\, e^{-2\pi T_2}\right)}
      -\mathcal{B}\, 
      e^{-b\,\left( T_1 + \mathcal{C}_2\, e^{-2\pi T_2}\right)}
      \\[3pt]
   \label{superpot_eff}
   &= \Bigl[ \mathcal{A}\, e^{-a\, T_1} -  
             \mathcal{B}\, e^{-b\, T_1}     \Bigr] 
      \;-\; 
      \Bigl[ \mathcal{A}\,\mathcal{C}_1\,a\, e^{-a\, T_1} -
             \mathcal{B}\,\mathcal{C}_2\,b\, e^{-b\, T_1}
      \Bigr]\, e^{-2\pi T_2} \;+ \ldots
\end{align}
with all K\"ahler moduli in Einstein frame. The constants 
$a=\frac{2\pi}{N}$ and $b=\frac{2\pi}{M}$ are determined by the rank of the
two gauge groups $U(N)$ respectively $U(M)$ and $\mathcal{A},\mathcal{B},\mathcal{C}_{(1,2)}$ 
are constant after 
the complex structure moduli have been stabilised 
at tree-level via \eqref{w_GVW}.

Next, we are going to analytically estimate the large volume minimum
of this scenario. 
However, we would like to emphasise that
the specific models 
of the following sections have been
analysed numerically 
for the K\"ahler potential \eqref{kahl-pot} and superpotential \eqref{instanton} without any approximations.
We start from the scalar F-term potential $V_F=e^{\mathcal{K}}\left( \lvert D W_{\rm np} \rvert^2-3\,\lvert W_{\rm np}\rvert^2\right)$ with $U_i$ and $S$ stabilised.
Since we are interested in a minimum at large $\mathcal{V}$, we expand this expression in powers of $1/\mathcal{V}$ and keep only the leading order term in $T_1$
\begin{equation}
  \label{rt_pot}
  V_F \sim \frac{ \sqrt{\tau_1}\:\bigl\vert \partial_{T_1} W_{\rm np} 
  \bigr\vert^2}{ \mathcal{V}}  
  + \mathcal{O}\Bigl(\mathcal{V}^{-2},e^{-4\pi\tau_2}\Bigr) \;.
\end{equation} 
The minimum of \eqref{rt_pot} is determined by $\partial_{T_1} W_{\rm np}=\partial_{\ov T_1} \ov W_{\rm np}=0$ stabilising $T_1$ at
\begin{equation}
  \label{fixed_T1}
  \tau^*_1 \simeq \frac{1}{a-b}\:
  \ln\,\biggl( \frac{\mathcal{A}\,a}{\mathcal{B}\, b} \biggr)
  \;,\qquad
  \rho^*_1=0\;,
\end{equation}
if $\mathcal{A}>\mathcal{B}$ are real and $a>b$. 
Using this solution, we identify the first term in \eqref{superpot_eff} as $W_0^{\rm eff}$ and in the second term we identify $A^{\rm eff}$ such that \eqref{instanton} becomes
\begin{equation}
  W^{\rm eff} = W_0^{\rm eff}
  - A^{\rm eff}\: e^{-2\pi T_2} + \ldots\;,
\end{equation}
which is of a form similar to \eqref{kklt}.
Since both $W_0^{\rm eff}$ and $A^{\rm eff}$  scale as $\exp\,(-a\tau_1^*)$, 
it is possible to obtain exponentially small values for them 
without fine-tuning.

We proceed and study the resulting effective potential for $\mathcal{V}$ and $T_2$ with $T_1$ stabilised at values \eqref{fixed_T1}. As for the LARGE Volume Scenario, we 
take the limit $\mathcal{V}\gg1$ and keep only the leading term in $\mathcal{V}$ 
at each order of $\exp\,(-2\pi \tau_2)$. The resulting potential then reads
(in Einstein frame)
\begin{align}
\label{effspot}
  V_F \sim& \hspace{10pt}
  \frac{8}{3}\, \Biggl( 
    \frac{\sqrt{\tau^*_1}}{\eta_1}\: \Bigl\lvert \gamma\, W_0^{\rm eff}
    \Bigr\rvert^2
    + (2\pi)^2\,\frac{\sqrt{\tau_2}}{\eta_1}\:
    \Bigl\lvert A^{\rm eff}\Bigr\rvert^2
    \Biggr)\,
    \frac{e^{-4\pi \tau_2}}{ \mathcal V}    \nonumber \\
  &-4\,\Bigl\lvert W_0^{\rm eff} \Bigr\rvert\, 
    \Bigl\lvert \tau_1^*\, \gamma\, W_0^{\rm eff} 
    + 2\pi \tau_2\, A^{\rm eff}
    \Bigr\rvert \:
    \frac{ e^{-2\pi \tau_2}}{\mathcal{V}^2} \\
  & +\frac{3\,\hat\xi}{4}\, 
    \Bigl\lvert W^{\rm eff}_0\Bigr\rvert^2\:
    \frac{1}{\mathcal{V}^3}\; ,
    \nonumber
\end{align}
with
$\gamma=(a\,\mathcal{C}_1-b\,\mathcal{C}_2)\left(\frac{1}{a}-\frac{1}{b}\right)^{-1}$
and the axion $\rho_2$ stabilised such that
the second term in \eqref{effspot} is minimised. 
The analytical analysis of this potential reveals
that $\mathcal{V}$ is fixed at 
\begin{equation}
\label{minV}  
  \mathcal{V}^* =  P\Bigl(A^{\rm eff},W_0^{\rm eff},\tau_1^*,\tau_2^*,
  \ldots\Bigr) \;
  e^{2\pi \tau_2^*}
\end{equation}
where $P$ is a rather complicated algebraic function of various quantities
and $\tau^*_2$ is determined by an implicit equation.
However, using the scaling $2\pi\tau_2\sim\ln\mathcal{V}$ obtained
from \eqref{minV}, 
we observe that those terms in \eqref{effspot} involving $\tau_1^*$ scale as $\mathcal{V}^{-3}$
and can therefore be absorbed into 
$\hat\xi$. The resulting potential is of a form as in the LVS and so we expect to find
a non-supersymmetric AdS minimum for large values of $\mathcal{V}$.


\bigskip
In order to obtain a positive cosmological constant, eventually the AdS minimum has to be 
uplifted. This is achieved for instance by anti $D3$-branes
at the bottom of a Klebanov-Strassler warped throat \cite{Klebanov:2000hb}. The corresponding uplift potential has the form
\begin{equation}
\label{uplift}
   V_{\rm up} \sim \frac{a^4}{\mathcal{V}^2}\: M_{\rm Pl}^4
   \qquad\mbox{and}\qquad
   a = e^{-\:\frac{2\pi  K}{3 g_s M}}
\end{equation}
is the warp factor at the bottom of the throat with 
$M,K$ the $A$- respectively $B$-cycle flux-quanta.


\section{Supersymmetric GUT Scenarios}
\label{sec_gut}

In the previous section, we have described a scenario 
featuring exponential control over $W^{\rm eff}_0$ and $A^{\rm eff}$ in an effective superpotential. 
As we will show in the following, by tuning the initial parameters (mostly 
at the order of 10\%), we are able
to dynamically fix the moduli such that $\mathcal{V}\sim 10^5$ and $\lvert W_0^{\rm eff}\rvert\sim 10^{-10}$. Expressing the string scale $M_{\rm s}=(\alpha')^{-1/2}$ and the gravitino mass $M_{3/2}$ in terms of $W_0^{\rm eff.}$ and $\mathcal{V}$ (in Einstein frame)
\begin{equation}
  \label{masses_01}
  M_{\rm s}= \frac{\sqrt{\pi}\, g_s^{1/4}}{\sqrt{\mathcal{V}}}
    \: M_{\rm Pl}
  \;,\hspace{40pt}
  M_{3/2}\sim \frac{\lvert W^{\rm eff}_0\rvert}
    {\mathcal{V}}\: M_{\rm Pl}\, ,
\end{equation}
we see that for these values we obtain $M_{\rm s}\simeq M_X\simeq 1.2\cdot 10^{16}\,{\rm GeV}$ and $M_{3/2}$ in the TeV range. Note that for the usual LVS, a high degree of tuning is needed to have
$M_{\rm s}\simeq M_X$ while keeping the susy breaking scale in the TeV regime.


\bigskip
Let us now investigate where in such a scenario the MSSM respectively the Grand Unified Theory might be localised. 
\begin{itemize}

\item A first guess is to wrap the MSSM  $D7$-branes on  
the four-cycle $\Gamma_{\rm b}$ with K\"ahler modulus $T_{\rm b}$
leading to a gauge coupling $\alpha^{-1}\simeq \tau_{\rm b}\sim \mathcal{V}^{2/3}$ which  is however too small. 

\item A second possibility is to place the branes on $\Gamma_2$ 
with K\"ahler modulus $T_2$ giving $\alpha^{-1}\simeq\tau_2 \simeq \frac{1}{2\pi} \ln\mathcal{V}\simeq 1.8$ which differs from the GUT
gauge coupling 
$\alpha^{-1}_X=25$ by an order of magnitude.

\item A third possibility is to wrap the $D7$-branes along $\Gamma_1$
with gauge coupling $\alpha^{-1}\simeq \tau_1\sim -\ln\,\lvert W_0^{\rm eff} \rvert\sim 21$ which is in the right ballpark.

\end{itemize}
Note  that we are certainly aware
of the problem raised in \cite{Blumenhagen:2007sm} about 
stabilising K\"ahler moduli via instantons if the MSSM or a GUT is realised by D-branes.
By considering a manifold with a further small K\"ahler modulus $T_4$ and wrapping the branes along $\Gamma_1+\Gamma_4$ with chiral intersection on $\Gamma_4$, we can avoid this issue.
The additional modulus then has to be stabilised by D-terms or, as suggested
in \cite{Cicoli:2008va}, by string loop-effects \cite{Berg:2007wt}.

\begin{figure}[t]
\begin{center} 
\includegraphics[width=0.5\textwidth]{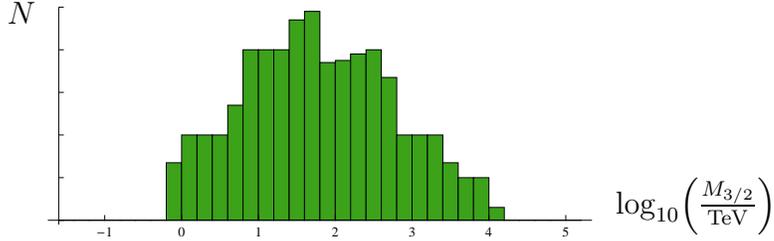}
\begin{picture}(0,0)
  \put(-225,82){$N$}
  \put(5,10){$\log_{10} \Bigl( \frac{M_{3/2}}{\rm TeV} \Bigr)$}
\end{picture}
\end{center} 
\vspace{-10pt}
\caption{{\small 
Distribution of $M_{3/2}$ with $\tau_1=25\pm 0.25$ 
and $\mathcal{V}/\sqrt{g_s}\simeq1.26\cdot10^5$ for a scan over natural values of $a,b,\mathcal{A},\mathcal{B}$. The statistical mean of this distribution is $\left\langle \log_{10}\left( M_{3/2}/{\rm TeV} \right)\right\rangle =1.79$ and the standard deviation is $\sigma = 0.98$.}\label{fighisto}}
\end{figure}

\bigskip
Having identified a natural choice for the MSSM  branes in our setup,
let us now take a different point of view and fix $\mathcal{V}/\sqrt{g_s}\simeq1.26\cdot10^5$ together with $\tau_1\simeq\alpha_{X}^{-1}\simeq25$. 
By scanning the parameters $a=\frac{2\pi}{N},b=\frac{2\pi}{M},\mathcal{A},\mathcal{B}$ in
a natural range, $M\in [2,12]$, $0<N<M$ and $\ln\mathcal{A}\in[\ln0.1,\ln10]$, 
$\ln\mathcal{B}\in[\ln\frac{\mathcal{A}}{10},\ln\mathcal{A})$
in equidistant steps,
we find the distribution of resulting values for $M_{3/2}$ shown in
figure \ref{fighisto}.
This illustrates that in our setup with input $M_{\rm s}=M_X$ and
$\alpha^{-1}=25$, a gravitino mass in the TeV region is obtained rather naturally.


\bigskip
We close the general analysis of our GUT setup with a summary of formulas for various mass scales: 
\begin{itemize}

\item Suppressing factors of $g_s$ and $2\pi$, the masses of the
closed sector moduli fields scale in the following way \cite{Conlon:2005ki}
\begin{equation}
\label{masses_02}
\arraycolsep1.5pt
\begin{array}{lcllcl}
  M_{U} &\sim& \displaystyle \frac{M_{\rm Pl}}{\mathcal{V} }\;,
    \hspace{70pt} &
  M_{T_1} &\sim&\displaystyle \lvert W^{\rm eff}_0\rvert \, 
    M_{\rm Pl}\;,
    \\[10pt]
  M_{T_2} &\sim&\displaystyle \frac{\lvert W^{\rm eff}_0\rvert}
    {\mathcal{V}}\: M_{\rm Pl}\;, &
  M_{T_{\rm b}} &\sim& \displaystyle \frac{\lvert W^{\rm eff}_0\rvert}
    {\mathcal{V}^{3/2}}\: M_{\rm Pl}\; ,
\end{array}
\end{equation}
and in the regime $W^{\rm eff}_0<{\cal V}^{-1}$ they are ordered from heavy to light.

\item For gravity mediated supersymmetry breaking, the gaugino masses
are calculated as\footnote{In the following, we can safely ignore
the effect of the uplifting \cite{Choi:2005ge}
 which gives only subdominant contributions
to the soft terms.}
\eq{
  \label{gaugino_grav}
  m^{\rm gravity}_{1/2}  = \frac{1}{2\,\mbox{Re}\,(f_a)} \:
    F^{I}   \partial_I\, \mbox{Re}\,(f_a) \;.
}

\item For anomaly mediated supersymmetry breaking, the following formula has to be 
evaluated \cite{Randall:1998uk,Bagger:1999rd} 
\eq{
  \label{gaugino_anom}
  m^{\rm anomaly}_{1/2,\, a}  & = -\:\frac{\alpha_a}{4\pi}\:
    \biggl( \bigl( 3 T_G-T_R\bigr)\:M_{3/2} + \bigl( T_G-T_R\bigr)\:
    K_I F^I \\
  & \hspace{158pt} + 
    \frac{2\, T_R}{d_R}\: \partial_I\bigl( \ln \det K\bigr\rvert_R''
    \bigr) F^I 
    \biggr) \;,
}
where a sum over all representations $R$ is understood. Furthermore, $T_G$ is the Dynkin index of the adjoint representation, $T_R$ is the Dynkin index of the representation $R$ and $d_R$ denotes its dimension. Finally, $K\rvert_R''$ is the K\"ahler metric restricted to the representation $R$ 
which for a {\em swiss-cheese} manifold takes the form \cite{Conlon:2006tj,Conlon:2006wz}
\eq{
  \bigl(K\bigr\rvert_R''\bigr)_{i\overline{j}} = 
    \frac{h_{i\overline j}\bigl(\tau_{\rm m},U\bigr)}{\tau_{\rm b}}
    \sim \frac{\tau_{\rm m}^{\lambda}}{\tau_{\rm b}}
    \:X_{i\overline j}\,\bigl(U\bigr) \;+\; 
    \ldots \;.
} 
Here, $\tau_{\rm m}$ denotes the volume of the small cycle the matter is localised on, in the present case $\tau_1$, and $\lambda$ can take values between $0$ and $1$. However, later we will use the value $\lambda=1/3$ of the minimal {\em swiss-cheese} setup \cite{Conlon:2006wz}. Finally, $X_{i\ov j}\,(U)$ is a matrix depending in the complex structure moduli.

\item The scalar masses obtained for gravity mediation of supersymmetry breaking read
\eq{
  \bigl( m_{0}^{\rm gravity} \bigr)^2 =
  M_{3/2}^2 + V_0 - F^I F^{\ov J}\partial_I\partial_J 
  \ln \tilde{K}_{\alpha}  \;,
}
where the potential in the minimum $V_0$ is set to zero and the K\"ahler potential for the matter fields is \cite{Conlon:2006tj,Conlon:2006wz}
\eq{
  \tilde{K}_{\alpha} = \frac{k_{\alpha}\bigl(\tau_{\rm m},U\bigr)}
  {\tau_{\rm b}}
  \sim \frac{\tau_{\rm m}^{\lambda}}{\tau_{\rm b}}\:
  k_{\alpha}^{(0)}\bigl(U\bigr) \;+\; \ldots \;.
}

\item Generically, the two-loop generated scalar masses for an 
anom\-a\-ly mediated scenario are 
always smaller than the supergravity mediated masses. Therefore, we do not display the relevant formulas here.

\end{itemize}


\subsection{Model 1: A Starter}
\label{sec_gut_1}

In the following subsections, we investigate the scalar F-term potential resulting
from the K\"ahler potential \eqref{kahl-pot} and the superpotential
\eqref{instanton} numerically, that is  we are searching for minima at large
volume with $\ln \mathcal{V}\sim2\,\pi \tau_2$.

\begin{figure}[h]
\centering
\renewcommand{\subfigcapskip}{-0pt}
\subfigure[$V_F(\mathcal{V},\tau_1)$]{
\includegraphics[width=0.3\textwidth]{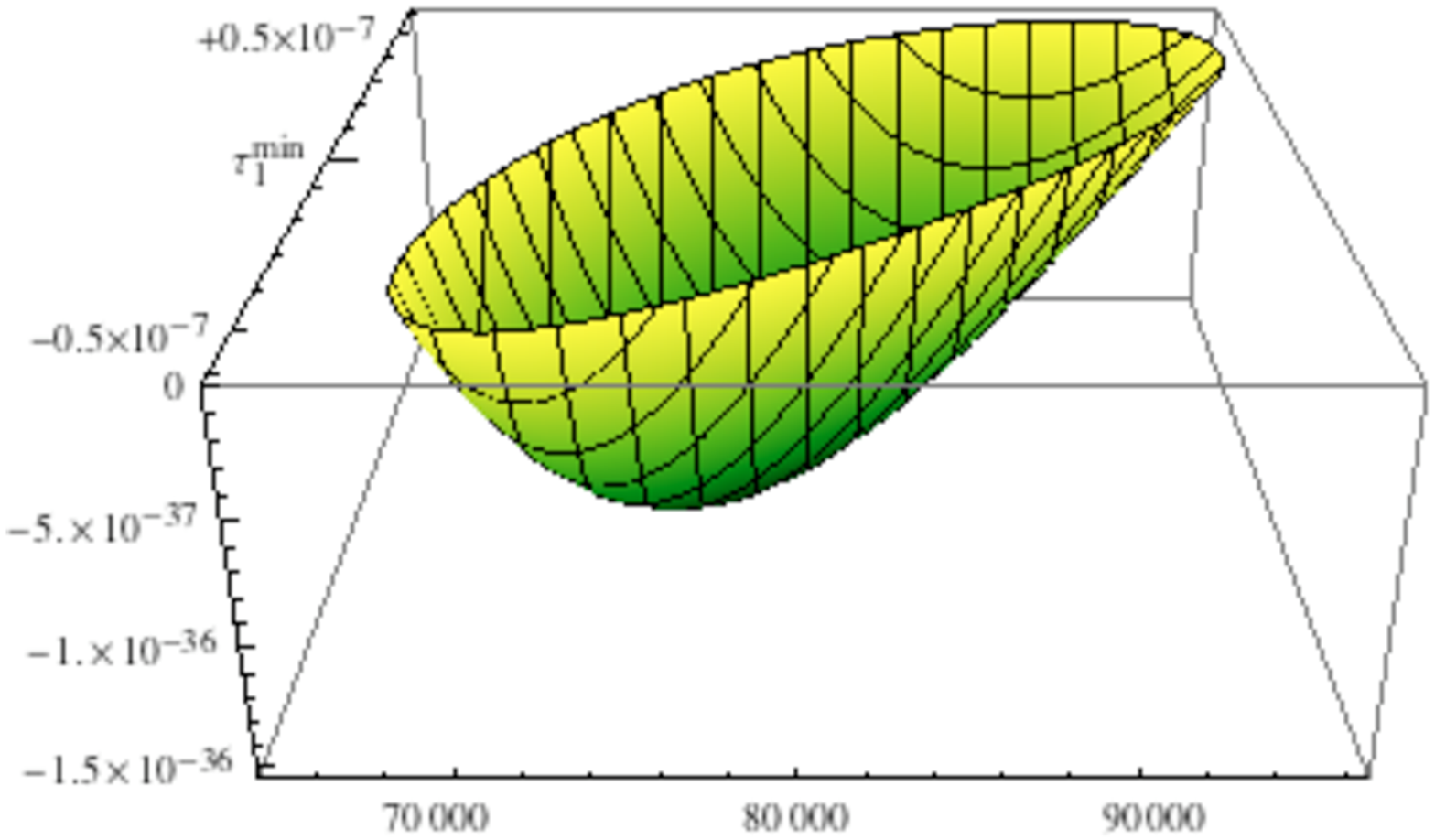}
}
\hfill
\subfigure[$V_F(\mathcal{V},\tau_2)$]{
\includegraphics[width=0.3\textwidth]{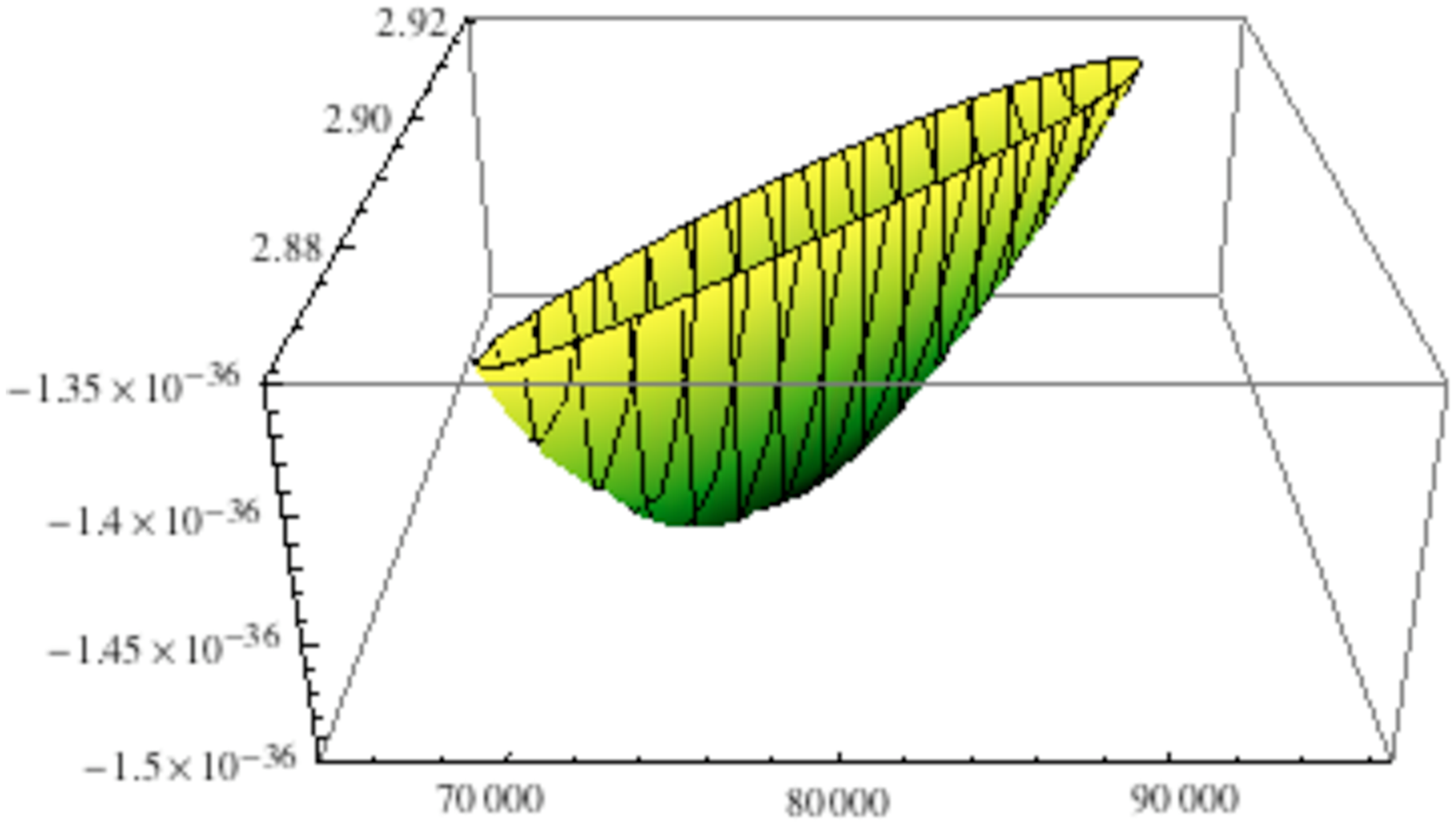} 
}
\hfill
\subfigure[$V_F(\tau_1,\tau_2)$]{
\includegraphics[width=0.3\textwidth]{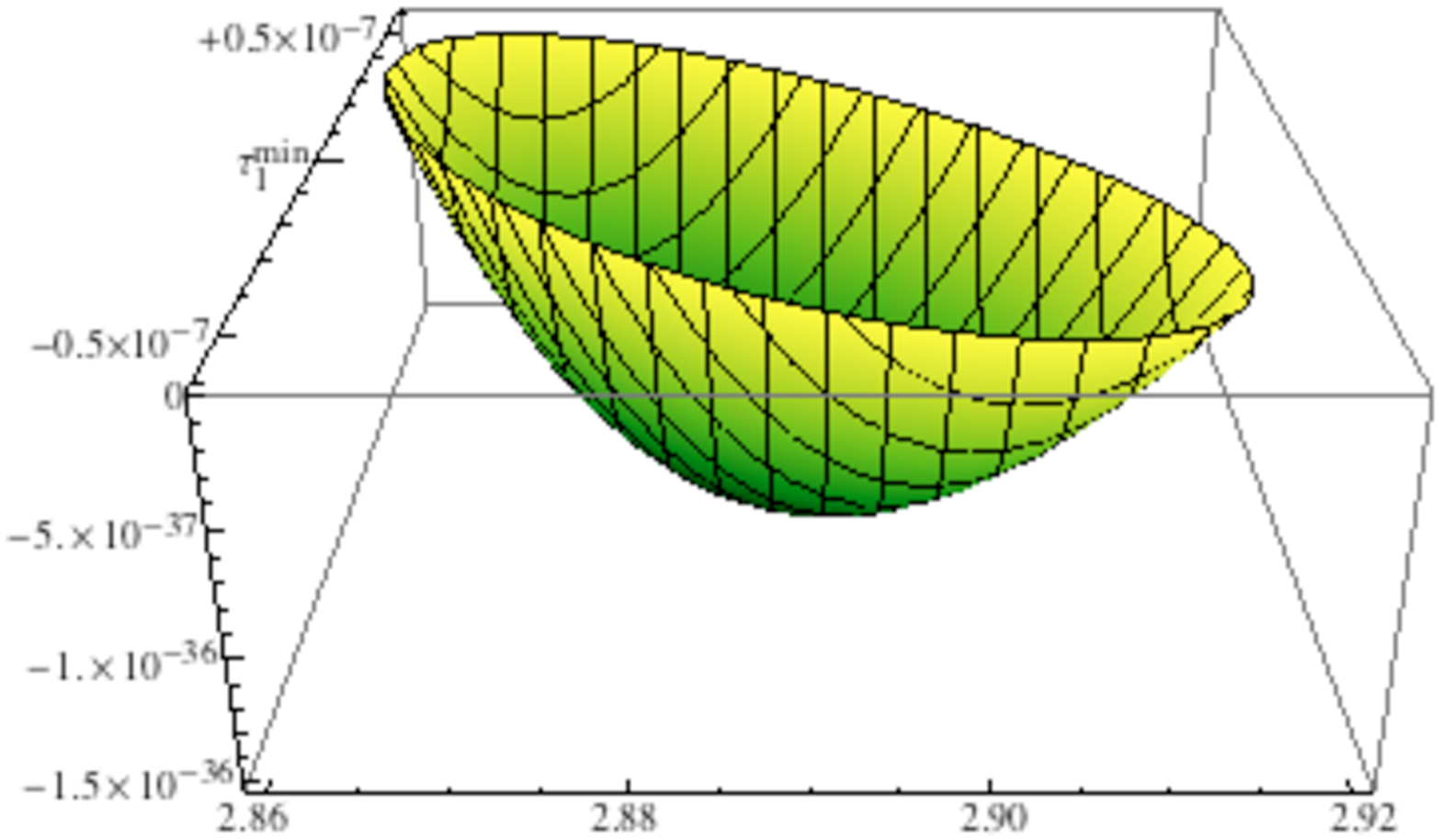} 
}
\vspace*{-7.5pt}
\caption{\small F-term potential of the GUT model 1
in the vicinity of the minimum. \label{figgut1}}
\end{figure}

\bigskip
Taking  the observations from the beginning of this section into account,
we first assume that the MSSM is localised on $D7$-branes
wrapping the cycle $\Gamma_1$ associated to the race-track modulus $T_1$ with 
size $\tau_1\simeq 25$.
 A particular set of parameters realising this setup without fine-tuning is the following
\begin{equation}
\label{parameters_naive}
\arraycolsep1.5pt
\begin{array}{@{}lcllcllcllcllcllcl@{}}
  \mathcal{A}  &=& 1.6\,,            \;\;&
  \mathcal{B}  &=& 0.2\,,            \;\;&    
  \mathcal{C}_1&=& 1\,,              \;\;&
  \mathcal{C}_2&=& 3\,,              \;\;&   
  a            &=& \frac{2\pi}{8}\,, \;\;&
  b            &=& \frac{2\pi}{9}\,,     \\[4pt]
  g_s          &=& \frac{2}{5}\,,    \;\;&   
  \eta_{\rm b} &=&  1\,,             \;\;&
  \eta_1       &=& \frac{1}{53}\,,   \;\;&
  \eta_2       &=& \frac{1}{6}\,,    \;\;&
  \chi         &=& \multicolumn{4}{l}{-136\,. }      
\end{array}
\end{equation}
We computed the scalar potential using the K\"ahler potential \eqref{kahl-pot}
and the  superpotential \eqref{instanton} without approximations and employed 
{\sf Mathematica} to determine the minimum with high precision. The resulting values are
\eq{
  \mathcal{V}^* = 78559\;,\quad
  T_1^* = 25.18\;,\quad
  T_2^* = 2.88\;,\quad
  V_F^* = -1.5\cdot 10^{-36}\, M_{\rm Pl}^4\;,
}
which is indeed the AdS large volume minimum argued
for in the last section. 
Three plots showing the potential in the vicinity of the 
minimum 
can be found in figures \ref{figgut1}.

\noindent
The various masses are computed using the formulas from above.
Let us however emphasise that for the gaugino and scalar masses it is crucial to work with precise 
values for $T_1^*$, $T_2^*$ and $\mathcal{V}^*$ in order to numerically obtain certain cancellations.
\begin{equation*}
\arraycolsep1.5pt
\renewcommand{\arraystretch}{1.25}
\begin{array}{@{\hspace{9pt}}lclll@{\hspace{9pt}}||@{\hspace{9pt}}
              lclll@{\hspace{9pt}}||@{\hspace{9pt}}
              lclll@{\hspace{9pt}}}
  \multicolumn{5}{@{\hspace{9pt}}c@{\hspace{9pt}}||@{\hspace{9pt}}}
    {\mbox{Fundamental Masses}} &
  \multicolumn{5}{c@{\hspace{9pt}}||@{\hspace{9pt}}}
    {\mbox{Moduli Masses}}      &
  \multicolumn{5}{c@{\hspace{9pt}}}{\mbox{Soft Masses}}        
    \\ \hline\hline
  M_{\rm s}                &=& 1.2\,\cdot&10^{16} &{\rm GeV} &
  M_{U}                    &=& 3.1\,\cdot&10^{13} &{\rm GeV} &
  m_{1/2}^{\rm gravity}      &=& 1.1\,\cdot&10^{-5} &{\rm GeV}       \\
  M_{3/2}                  &=& 1.6\,\cdot&10^4    &{\rm GeV} &
  M_{T_1}                  &=& 1.2\,\cdot&10^{9}  &{\rm GeV} &
  m_{1/2,\, \scriptscriptstyle SU(3)}^{\rm anomaly} 
                           &=& 1.2\,\cdot&10^{-3} &{\rm GeV}       \\
  &&&&&
  M_{T_2}                  &=& 1.6\,\cdot&10^{4}  &{\rm GeV} &
  m_{0}^{\rm gravity} &=& 64&           &{\rm GeV}            \\
  &&&&&
  M_{T_{\rm b}}            &=& 56 &               &{\rm GeV}             
\end{array}
\end{equation*}

\pagebreak[1]
\noindent
Let us comment on these scales:
\vspace*{-1.75pt}
\begin{itemize}
\setlength{\itemsep}{0pt}

\item By construction, the string scale $M_{\rm s}$ coincides with the GUT scale and the gravitino mass $M_{3/2}$ is in the TeV regime.

\item The closed sector moduli masses are ordered from heavy to light and take acceptable values except for $T_{\rm b}$ which is too small. 
For $M_s=M_X$, the lower bound for moduli masses not to be in
conflict with cosmology 
is around $1\,{\rm TeV}$.  Therefore, in this model we face the 
Cosmological Moduli Problem (CMP) \cite{Banks:1993en,deCarlos:1993jw}.

\item In addition, the gaugino as well as the scalar masses are far too small. The main reason is that we realised the MSSM on the cycle $\Gamma_1$ which is related to the race-track modulus $T_1$.
For this modulus we observe numerical cancellations giving
$F^1\simeq 2\cdot 10^{-22}\, M_{\rm Pl}\sim 10^{-8}\, M_{3/2}$, i.e.
supersymmetry breaking on the race-track cycle is eight orders of magnitude
smaller than naively expected.
The explanation is that the exact race-track minimum, in leading order
given by the globally supersymmetric minimum \eqref{fixed_T1}, 
is almost supersymmetric.
For the gravity mediated gaugino masses we therefore obtain a strong suppression
\eq{
  \label{soft_scaling_1}
   m_{1/2}^{\rm gravity}=\frac{1}{2\,\tau_1}\: 
   F^1\sim \frac{1}{2\cdot 25} \:2\cdot 10^{-22} \:M_{\rm Pl}
   \sim 10^{-5}\, {\rm GeV} \;.
}   

\item For the anomaly mediated gaugino masses
$m_{1/2}^{\rm anomaly}$, we find the expected cancellation of $M_{3/2}$ at leading order (see \cite{Conlon:2006wz} for a detailed derivation), however the sub-leading order in $\mathcal{V}$ dominates over $F^{1}$
\eq{
   \label{soft_scaling_2}
   m_{1/2,\, \scriptscriptstyle SU(3)}^{\rm anomaly} & \sim 
   \frac{\alpha_a}{4\pi}\:\biggl( 3\,T_G\,M_{3/2}\, \Bigl( 1-1+
   \mathcal{O}\bigl(\mathcal{V}^{-1}\bigr) \Bigr)
   +2\,\lambda\, T_R\: \frac{F^1}{2\,\tau_1} \biggr) \\
   &\sim \frac{1}{300}\:\biggl(3\cdot3\cdot 10^4\,{\rm GeV}\cdot10^{-5}
     + 2\, \lambda\cdot6\cdot \frac{2\cdot 10^{-22}\,M_{\rm Pl}}{2\cdot 25}
     \biggr) \\[3pt]
   & \sim 10^{-3}\,{\rm GeV} \;.
}
This value is of course 
still too small but note it is larger than the gravity mediated term.
The gaugino masses are thus dominantly generated via anomaly mediation.

\item A similar mechanism is at work for the scalar masses where $M_{3/2}^2$
is cancelled at leading order and subleading corrections in $\mathcal{V}$
give the main contribution 
(see again \cite{Conlon:2006wz} for a detailed derivation of this expression)
\eq{
  \label{soft_scaling_3}
  \bigl( m_{0}^{\rm gravity}\bigr)^2 &\sim
  M_{3/2}^2\,\Bigl( 1-1 + \mathcal{O}\bigl( \mathcal{V}^{-1}\bigr) \Bigr)
  +\lambda\,\biggl( \frac{F^1}{2\,\tau_1} \biggr)^2 \\
  &\sim \bigl( 10^4\, {\rm GeV} \bigr)^2\cdot 10^{-5}
  +\lambda\, \biggl( \frac{2\cdot 10^{-22}\, M_{\rm Pl}}{2\cdot 25} 
  \biggr)^2 \\[3pt]
  &\sim \bigl( 10^{3/2}\,{\rm GeV} \bigr)^2 \;.
}

\end{itemize}

In conclusion, although we were able to easily arrange for $M_{\rm s}\simeq M_X$, $\alpha^{-1}\simeq\alpha^{-1}_X\simeq25$ and a gravitino mass in the TeV range, the soft terms are much too small. 
In order to obtain realistic soft masses, 
two options seem viable. 
Either we take the present setup and
scale the mass parameters by a factor of $10^6$, or we wrap the MSSM branes
not only along $\Gamma_1$ but on $\Gamma_1+\Gamma_2$ giving also a contribution from $F^2$ to the soft terms.
In the following two subsections, we discuss these
two possibilities in more detail.


\subsection[Model 2: A Mixed Anomaly-Supergravity Mediated Model]{Model 2: A Mixed Anomaly-Gravity Mediated Model}

\begin{figure}[p]
\centering
\renewcommand{\subfigcapskip}{-0pt}
\subfigure[$V_F(\mathcal{V},\tau_1)$]{
\includegraphics[width=0.3\textwidth]{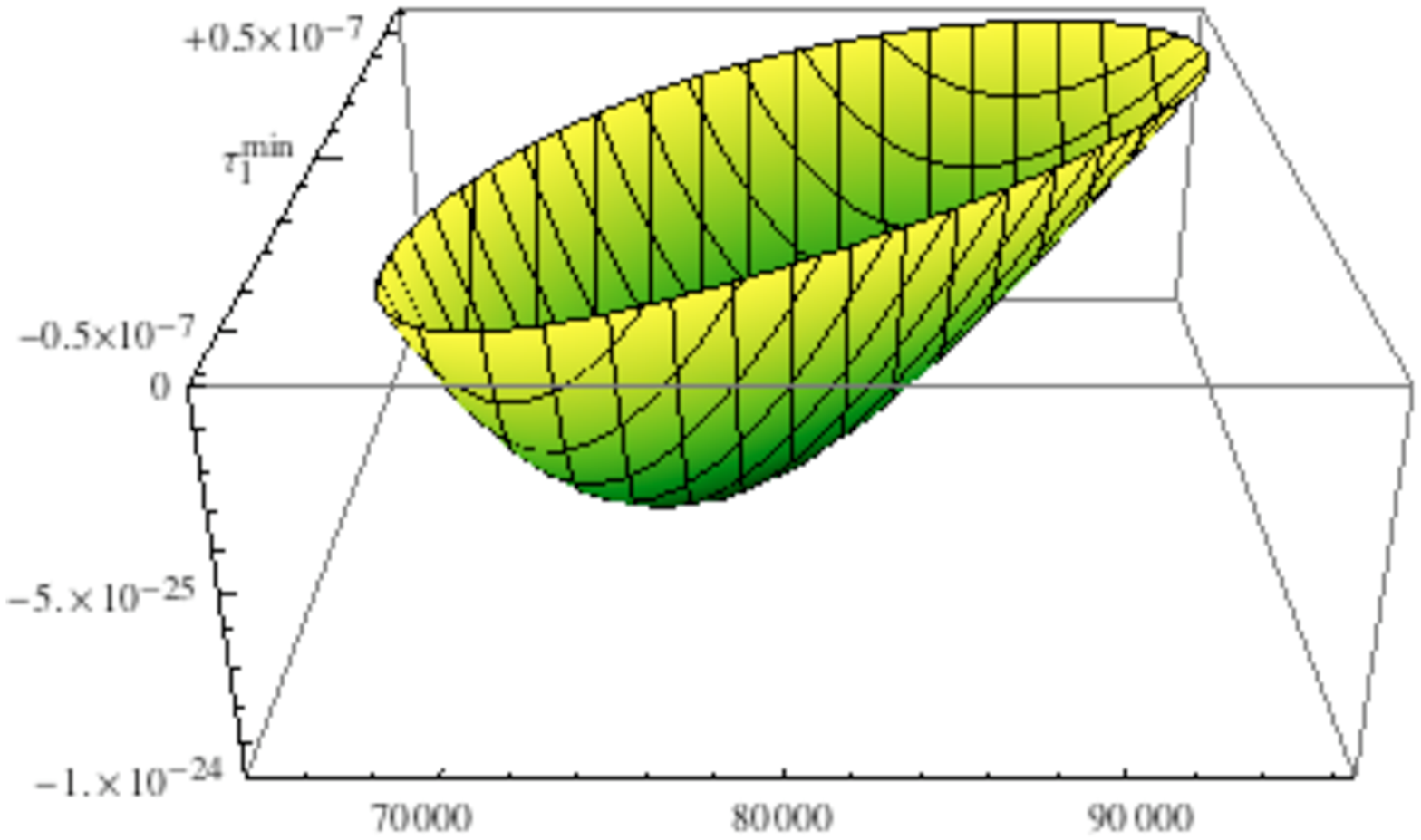}
}
\hfill
\subfigure[$V_F(\mathcal{V},\tau_2)$]{
\includegraphics[width=0.3\textwidth]{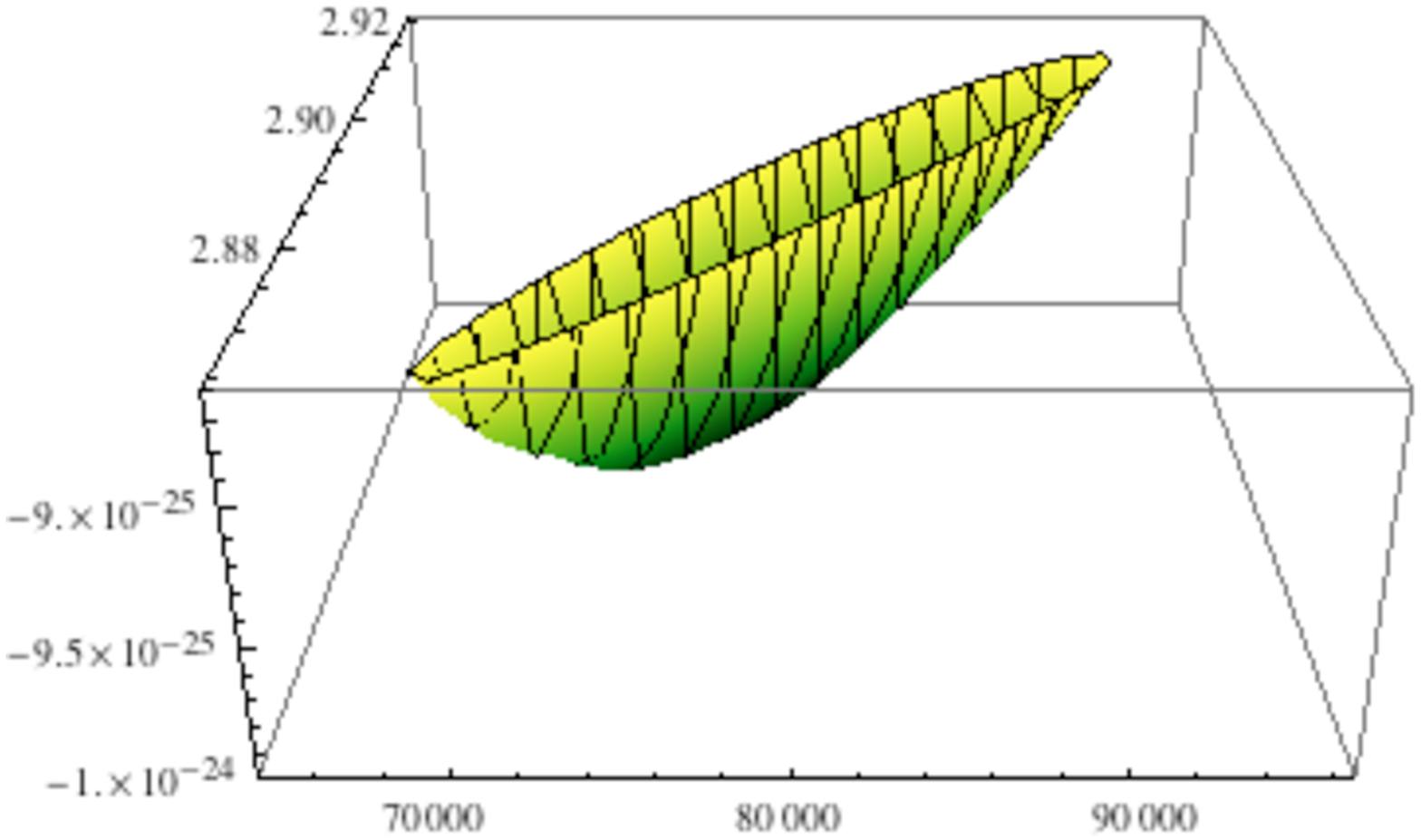} 
}
\hfill
\subfigure[$V_F(\tau_1,\tau_2)$]{
\includegraphics[width=0.3\textwidth]{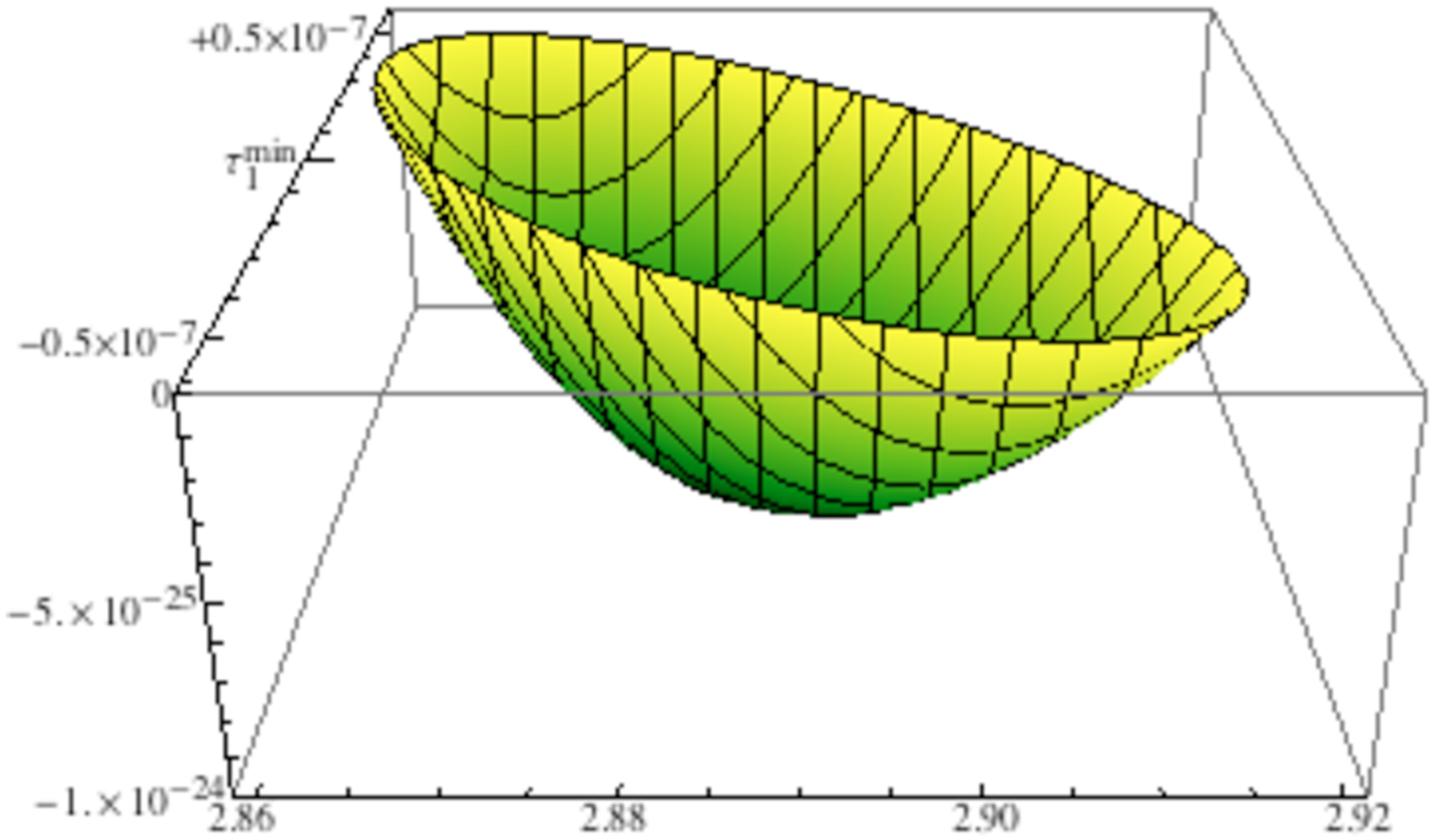} 
}
\vspace*{-5pt}
\caption{\small F-term potential of the GUT model 2
in the vicinity of the minimum. \label{figgut2}}
\vspace{-11pt}
\end{figure}

Recall that the minimum for the race-track modulus $T_1$ is approximately 
supersymmetric. To construct a model with gaugino masses in the TeV range, 
we use the same set of parameters \eqref{parameters_naive} of the previous setup but scale $\mathcal{A}$ and $\mathcal{B}$ to somewhat more unrealistic values
\eq{
  \label{model2_scaling}
  \mathcal{A}=1.6\;\times\;8\cdot10^5\;,\qquad
  \mathcal{B}=0.2\;\times\;8\cdot10^5\;.
}
The minimum of the resulting F-term potential is not changed except for the value of $V_F$ in the minimum
\eq{
  \mathcal{V}^* = 78559\;,\quad
  T_1^* = 25.18\;,\quad
  T_2^* = 2.88\;,\quad
  V_F^* = -9.5\cdot 10^{-25}\, M_{\rm Pl}^4\;.
}
Three plots showing the potential in the vicinity of the minimum can be found in figures \ref{figgut2} on page \pageref{figgut2} and the  mass scales for the new setup are the following:
\begin{equation*}
\arraycolsep1.5pt
\renewcommand{\arraystretch}{1.25}
\begin{array}{@{\hspace{9pt}}lclll@{\hspace{9pt}}||@{\hspace{9pt}}
              lclll@{\hspace{9pt}}||@{\hspace{9pt}}
              lclll@{\hspace{9pt}}}
  \multicolumn{5}{@{\hspace{9pt}}c@{\hspace{9pt}}||@{\hspace{9pt}}}
    {\mbox{Fundamental Masses}} &
  \multicolumn{5}{c@{\hspace{9pt}}||@{\hspace{9pt}}}
    {\mbox{Moduli Masses}}      &
  \multicolumn{5}{c@{\hspace{9pt}}}{\mbox{Soft Masses}}        
    \\ \hline\hline
  M_{\rm s}                &=& 1.2\,\cdot&10^{16} &{\rm GeV} &
  M_{U}                    &=& 3.1\,\cdot&10^{13} &{\rm GeV} &
  m_{1/2}^{\rm gravity}      &=& 8.6       &        &{\rm GeV}       \\
  M_{3/2}                  &=& 1.3\,\cdot&10^{10} &{\rm GeV} &
  M_{T_1}                  &=& 9.9\,\cdot&10^{14} &{\rm GeV} &
  m_{1/2,\, \scriptscriptstyle SU(3)}^{\rm anomaly} 
                           &=& 962       &        &{\rm GeV}       \\
  &&&&&
  M_{T_2}                  &=& 1.3\,\cdot&10^{10} &{\rm GeV} &
  m_{0}^{\rm gravity} &=& 5.1\,\cdot&10^7    &{\rm GeV}       \\
  &&&&&
  M_{T_{\rm b}}            &=& 4.5\,\cdot&10^7    &{\rm GeV}            
\end{array}
\end{equation*}

\pagebreak[2]
\noindent
We again comment on these scales:
\begin{itemize}

\item Since the value of the volume modulus is not changed compared to the previous setup, we similarly obtain $M_{\rm s}\simeq M_X$. However, the gravitino mass is in an intermediate regime due to the change in $W_0^{\rm eff}$. 

\item With the gravitino masses in the intermediate regime,  
the Cosmological Moduli Problem has been evaded, 
as $T_{\rm b}$ is  much heavier then the TeV scale. 
We also see that $M_{T_1}>M_U$, but since we have a well-defined expansion in $1/\mathcal{V}$ we can safely stabilise $U$ and then study the stabilisation of $T_1$.

\item Concerning the gravity mediated gaugino mass, the scaling \eqref{model2_scaling} results in a value of $W_0^{\rm eff}$ which is $8\cdot 10^5$ times larger than in the previous setup leading to 
$F^1\sim 10^{-16}\, M_{\rm Pl}$. The gravity mediated gaugino mass $m_{1/2}^{\rm gravity}$ is still too small but the anomaly mediated masses   are now in the  TeV regime.

\item As expected from the first example, the gravity mediated scalar masses 
are at $m^{\rm gravity}_0\sim 10^7\,{\rm GeV}$ and therefore much heavier than the
gaugino masses. However, they come in complete $SU(5)$ multiplets
and therefore do not spoil gauge coupling unification.

\item With this hierarchy between the gaugino masses and the scalar
masses, we have a dynamical realisation of the split supersymmetry
scenario \cite{ArkaniHamed:2004fb}. The Higgs sector masses, i.e. the canonically normalised
$\hat\mu$-term and the soft term $\mu B$, are expected to be of the same order of magnitude as  the scalar masses, simply
for the reason that here both $F^{\rm b}\sim M_{3/2}$ and $F^1$ contribute.
Therefore, in order to keep these at the weak scale,
a fine-tuning of the supersymmetric $\mu$-term is necessary.

\end{itemize}

To summarise, tuning the initial parameters such that
 $\alpha^{-1}\simeq\alpha_X^{-1}\simeq 25$ and $M_s=M_X$, and localising
the MSSM solely on the race-track cycle $\Gamma_1$,
we find a high suppression of the gravity mediated
gaugino masses due to the
quasi-supersymmetry of the race-track minimum. 
Fixing the gaugino masses at the
TeV scale  leads to an intermediate supersymmetry
breaking scenario with gravitino masses and
scalar masses in the intermediate regime.
Due to the large value of $M_{3/2}$, this evades
the CMP and gives a stringy realisation of
the split supersymmetry scenario proposed
in \cite{ArkaniHamed:2004fb}.\footnote{For a local realisation of split supersymmetry see \cite{Antoniadis:2004dt}, and a realisation by mixed
anomaly-D-term mediation has been reported in \cite{Dudas:2005vv}.}


\subsection{Model 3: An LVS like Supergravity Mediated Model}
\label{sec_gut_3}

We now consider the second possibility from the end of section \ref{sec_gut_1} which indeed realises 
our initial goal, namely to naturally find  large volume
minima with the string scale at the GUT scale, gauge coupling
unification  and soft
masses in the TeV regime. This provides a concrete moduli
stabilisation scenario for which the analysis of
\cite{Aparicio:2008wh} is applicable. There, the computation
of soft masses and running to the weak scale has 
been studied quite systematically for moduli dominated
supersymmetry breaking in F-theory respectively Type IIB
orientifold compactifications realising the 
MSSM.

\begin{figure}[p]
\centering
\renewcommand{\subfigcapskip}{-0pt}
\subfigure[$V_F(\mathcal{V},\tau_1)$]{
\includegraphics[width=0.3\textwidth]{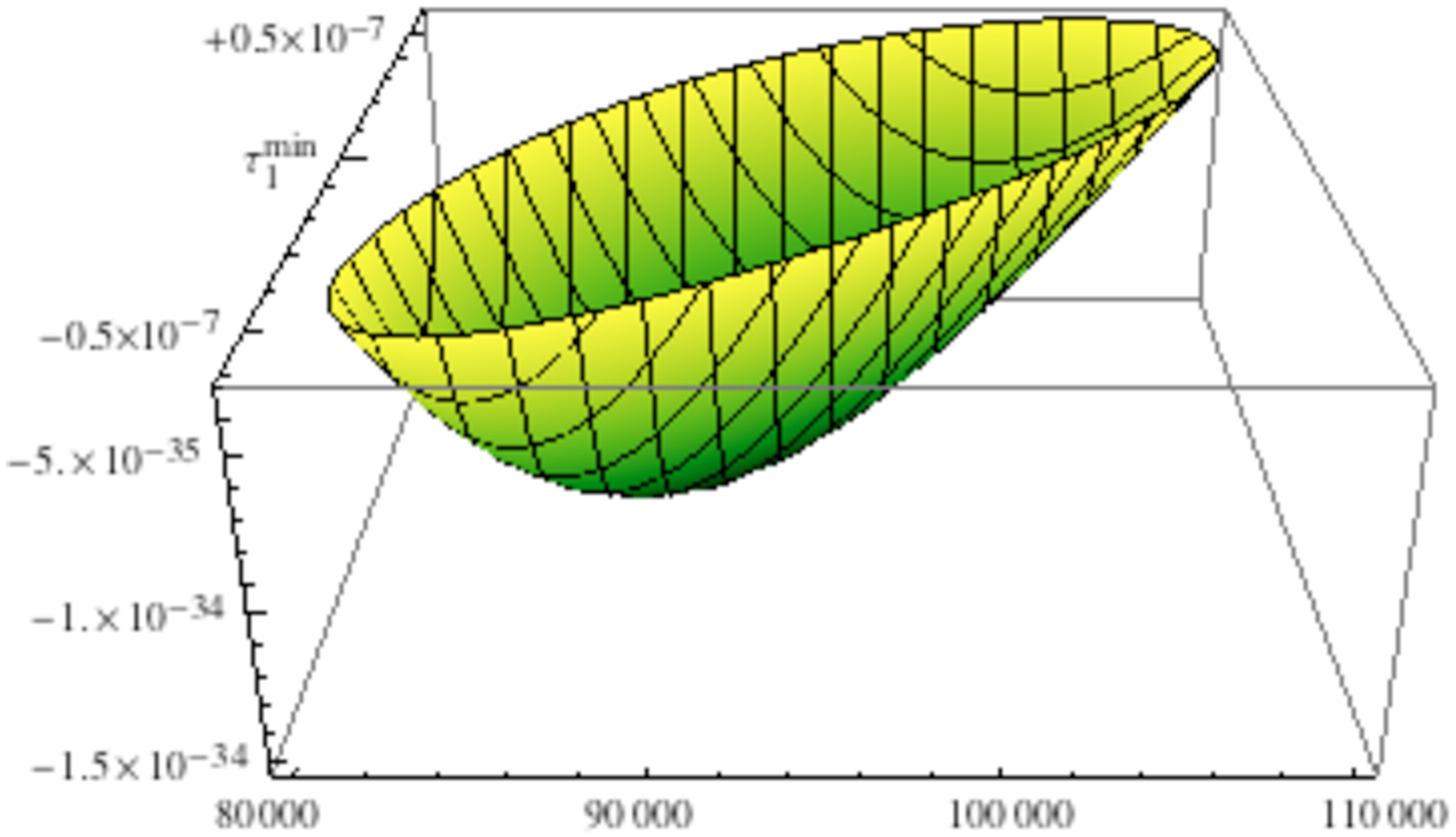}
}
\hfill
\subfigure[$V_F(\mathcal{V},\tau_2)$]{
\includegraphics[width=0.3\textwidth]{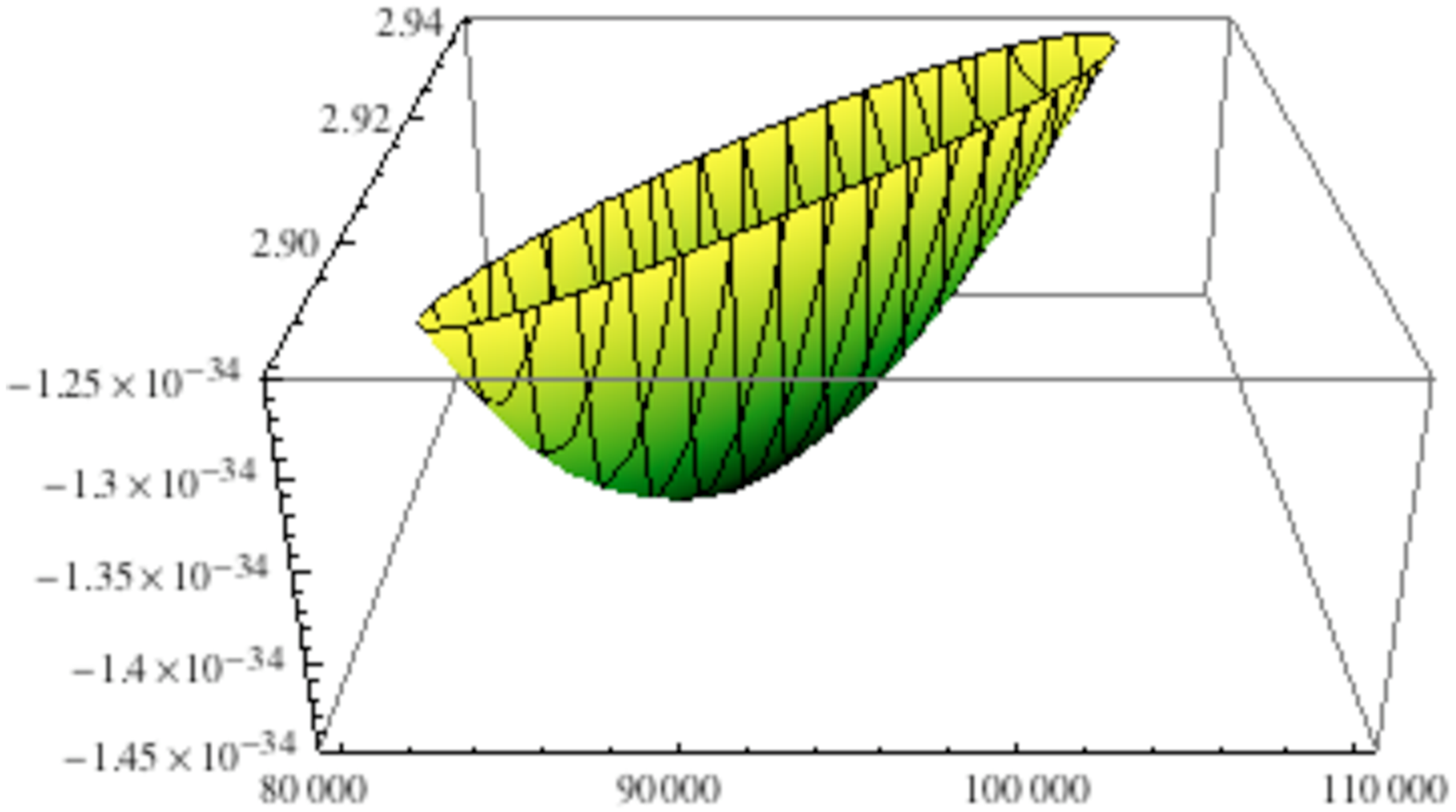} 
}
\hfill
\subfigure[$V_F(\tau_1,\tau_2)$]{
\includegraphics[width=0.3\textwidth]{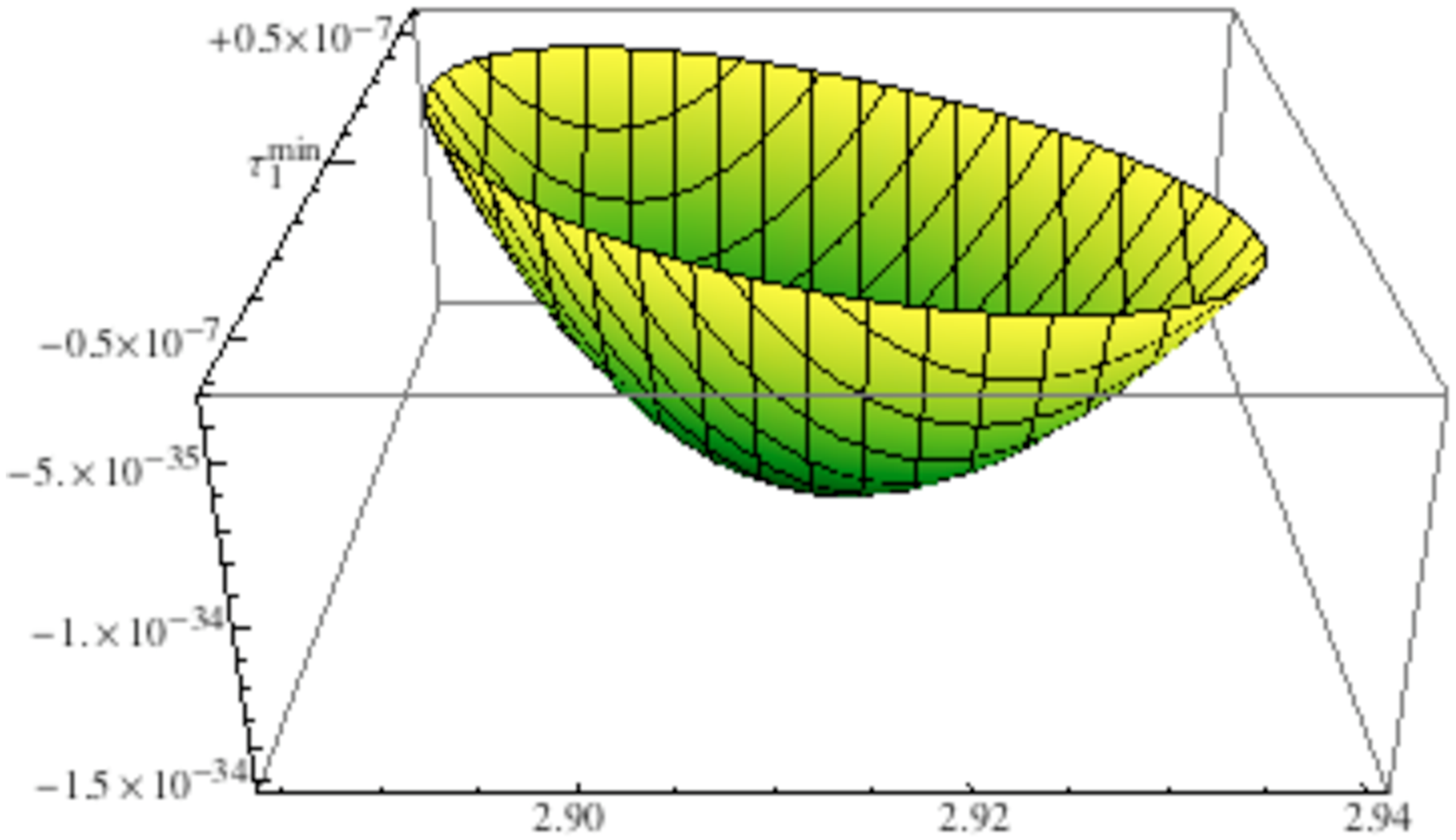} 
}
\vspace*{-5pt}
\caption{\small F-term potential of the GUT model 3
in the vicinity of the minimum. \label{figgut3}}
\vspace{-11pt}
\end{figure}

\bigskip 
We modify our original setup such that the soft terms are dominantly generated via $T_2$ similarly to the original LVS. To do so, we place the MSSM  $D7$-branes on the combination of cycles $\Gamma_1+\Gamma_2$. A concrete set of parameters realising this setup without fine-tuning is
\begin{equation}
\arraycolsep1.5pt
\begin{array}{@{}lcllcllcllcllcllcl@{}}
  \mathcal{A}  &=& 1.5\,,            \;\;&
  \mathcal{B}  &=& 0.25\,,            \;\;&    
  \mathcal{C}_1&=& 1\,,              \;\;&
  \mathcal{C}_2&=& 3\,,              \;\;&   
  a            &=& \frac{2\pi}{8}\,, \;\;&
  b            &=& \frac{2\pi}{9}\,,     \\[4pt]
  g_s          &=& \frac{2}{5}\,,    \;\;&   
  \eta_{\rm b} &=&  1\,,             \;\;&
  \eta_1       &=& \frac{1}{40}\,,   \;\;&
  \eta_2       &=& \frac{1}{6}\,,    \;\;&
  \chi         &=& \multicolumn{4}{l}{-153\,. }      
\end{array}
\end{equation}
The minimum of the scalar F-term potential is again determined with the help of {\sf Mathematica} giving
\eq{
  \mathcal{V}^* = 92158\;,\quad
  T_1^* = 21.58\;,\quad
  T_2^* = 2.91\;,\quad
  V_F^*= -1.4\cdot 10^{-34}\, M_{\rm Pl}^4\;,
}
so that the gauge coupling of the D7-branes $\alpha^{-1}=\tau_1+\tau_2=24.8$ is again the unified gauge coupling at the GUT scale. Three plots showing the potential in the vicinity of the minimum can be found in figures \ref{figgut3} on page \pageref{figgut3}.
The mass scales in this setup are calculated as follows:
\begin{equation*}
\arraycolsep1.5pt
\renewcommand{\arraystretch}{1.25}
\begin{array}{@{\hspace{9pt}}lclll@{\hspace{9pt}}||@{\hspace{9pt}}
              lclll@{\hspace{9pt}}||@{\hspace{9pt}}
              lclll@{\hspace{9pt}}}
  \multicolumn{5}{@{\hspace{9pt}}c@{\hspace{9pt}}||@{\hspace{9pt}}}
    {\mbox{Fundamental Masses}} &
  \multicolumn{5}{c@{\hspace{9pt}}||@{\hspace{9pt}}}
    {\mbox{Moduli Masses}}      &
  \multicolumn{5}{c@{\hspace{9pt}}}{\mbox{Soft Masses}}        
    \\ \hline\hline
  M_{\rm s}                &=& 1.1\,\cdot&10^{16} &{\rm GeV} &
  M_{U}                    &=& 2.6\,\cdot&10^{13} &{\rm GeV} &
  m_{1/2}^{\rm gravity}      &=& 819       &        &{\rm GeV}       \\
  M_{3/2}                  &=& 168       &        &{\rm TeV} &
  M_{T_1}                  &=& 1.5\,\cdot&10^{10} &{\rm GeV} &
  m_{1/2,\,\scriptscriptstyle SU(3)}^{\rm anomaly} 
                           &=& 10        &        &{\rm GeV}       \\
  &&&&&
  M_{T_2}                  &=& 168       &        &{\rm TeV} &
  m_{0}^{\rm gravity} &=& 817       &  \,      &{\rm GeV}       \\
  &&&&&
  M_{T_{\rm b}}            &=& 553      &         &{\rm GeV}            
\end{array}
\end{equation*}

\pagebreak[1]
\noindent
Let us again comment on the various scales arising
in this large volume minimum:
\begin{itemize}

\item We arranged the parameters for the present setup such that $M_{\rm s}\simeq M_X$ together with the gravitino mass in the TeV range.

\item The closed sector moduli masses take (almost) acceptable values with \linebreak[4]$M_{T_{\rm b}}$ 
close to  the  regime where the Cosmological Moduli Problem may be avoided. 
(Note that we were not careful with factors of $2\pi$.)

\item The soft masses are dominantly generated via the supersymmetry breaking of $T_2$ specified by $F^2\sim10^{-14} M_{\rm Pl}$. 
Since $T_2$ is stabilised as in the original LARGE Volume Scenario, similar mechanisms generating the soft masses are at work \cite{Conlon:2006wz}. In particular, using the fact that $F^1\sim 10^{-20}M_{\rm Pl}\ll F^2$, 
the common term determining the gaugino as well as the scalar masses can be expressed as
\eq{
  \frac{F^1+F^2}{2\,\bigl( \tau_1+\tau_2\bigr)} \sim
  \frac{F^2}{2\,\bigl( \tau_1+\tau_2\bigr)} \sim
  \frac{\tau_2}{\tau_1+\tau_2}\: 
  \frac{M_{3/2}}{\ln\left( M_{\rm Pl}/M_{3/2}\right)} \;.
} 
Here we used that $F^2\simeq  2 \tau_2\,M_{3/2}/\ln\left( M_{\rm Pl}/M_{3/2}\right)$ which was obtained in \cite{Conlon:2006us}. For the gravity mediated gaugino mass we thus find
\eq{
   m_{1/2}^{\rm gravity}=\frac{F^1+F^2}{2\,\bigl(\tau_1+\tau_2\bigr)}\: 
   \sim \frac{3}{25} \:\frac{1.7\cdot 10^{5}\,{\rm GeV}}
   {\ln\bigl( 10^{18}\cdot 10^{-5}\bigr)}
   \sim 700\, {\rm GeV} \;.
}

\item For the anomaly mediated gaugino mass we use equation \eqref{soft_scaling_2} to obtain
\begin{align}
  \nonumber
   m_{1/2,\, \scriptscriptstyle SU(3)}^{\rm anomaly} & \sim 
   \frac{\alpha_a}{4\pi}\:\biggl( 3\,T_G\,M_{3/2}\, \Bigl( 1-1+
   \mathcal{O}\bigl(\mathcal{V}^{-1}\bigr) \Bigr)
   +2\,\lambda\, T_R\: \frac{F^1+F^2}{2\,\bigl(\tau_1+\tau_2\bigr)}
    \biggr) \\
   &\sim \frac{1}{300}\:\biggl(3\cdot3\cdot 10^5\,{\rm GeV}\cdot10^{-6}
     + 2\, \lambda\cdot6\cdot 700\, {\rm GeV}
     \biggr) \\[3pt]
  \nonumber 
   & \sim 10\,{\rm GeV} \;,
\end{align}
with $\lambda=1/3$ \cite{Conlon:2006wz}. Note that again the contribution of $M_{3/2}$ is cancelled at leading order but the subleading correction $M_{3/2}/\mathcal{V}$ is suppressed compared to $F^2$.

Furthermore, in this supersymmetry breaking scheme, the anomaly mediated 
gaugino masses are significantly smaller than the gravity mediated ones.
This is contrast to the so-called mirage mediation scheme
arising for instance in the original KKLT scenario 
\cite{Endo:2005uy,Choi:2005uz,Choi:2007ka},
where both are of the same order of magnitude.

\item Finally, we consider the gravity mediated scalar masses. Referring to formula \eqref{soft_scaling_3}, we obtain
\eq{
  \bigl( m_{0}^{\rm gravity}\bigr)^2 &\sim
  M_{3/2}^2\,\Bigl( 1-1 + \mathcal{O}\bigl( \mathcal{V}^{-1}\bigr) \Bigr)
  +\lambda\,\biggl( \frac{F^1+F^2}{2\,\bigl(\tau_1+\tau_2\bigr)} 
  \biggr)^2 \\[3pt]
  &\sim \bigl( 10^5\, {\rm GeV} \bigr)^2\cdot 10^{-6}
  +\lambda\, \bigl( 700\,{\rm GeV}  \bigr)^2 \\[4pt]
  &\sim \bigl( 10^{2}\,{\rm GeV} \bigr)^2 \;.
}
Note that for the scalar masses, the contribution from  $M_{3/2}$ is cancelled at leading order but now the subleading corrections scale as $M_{3/2}/\sqrt{\mathcal{V}}$. Therefore, by accident, the two terms in the equation above are of the same order which is in contrast to the relation $m_{0}^{\rm gravity}= m_{1/2}^{\rm gravity}/\sqrt{3}$ obtained in the LVS \cite{Conlon:2006wz}.

\item Assuming that  the supersymmetric $\mu$-term
vanishes, the canonical normalised Higgs parameters are also in the
TeV regime. This would  solve the $\mu$-problem via
the Giudice-Masiero mechanism \cite{Giudice:1988yz}.

\end{itemize}

In summary, by wrapping the MSSM supporting $D7$-branes along $\Gamma_1+\Gamma_2$,
we obtain a LARGE Volume Scenario with supergravity mediated
soft masses in the TeV region and $M_{\rm s}=M_X$. This is different compared to the original LVS
where the string scale is usually at an  intermediate scale.
Along the lines of \cite{Conlon:2007xv,Aparicio:2008wh}, the next step is 
to calculate the soft-terms at the weak scale to see whether
distinctive patterns for the supersymmetric
phenomenology to be tested at the LHC can be obtained.


\section{Comment on the Cosmological Constant}
\label{sec_cc}

For the GUT setup discussed in the previous section, the tree-level
cosmological constant is 
$V_F^*=-1.4\cdot 10^{-34}M_{\rm Pl}^4$
and therefore a high degree of fine-tuning in the uplift potential
\eqref{uplift} is needed to obtain
$\Lambda \simeq +10^{-120}\,M_{\rm Pl}^4$.
After such a fine-tuning has been achieved, a LARGE volume scenario
with $M_{\rm s}\simeq M_X$ contains
a natural candidate serving as a quintessence field \cite{Choi:1999xn,Svrcek:2006hf,Quevedo:talksp08}.
Indeed, taking into account also non-perturbative corrections corresponding
to the large four-cycle $\Gamma_{\rm b}$, the scalar potential
depending on the (canonically normalised) axion $\sigma_{\rm b}$ takes the form
\eq{
  \label{quintessence}
  \frac{V_Q}{M_{\rm Pl}^4}\simeq \left( \frac{M_{\rm Pl}\, 
  M_{3/2}}{M_X^2}\right)\,
  \:e^{-\frac{2\pi}{L} \tau_{\rm b}}\: \biggl( 1-\cos\biggl(
  \frac{2\pi}{L}\, \sigma_{\rm b} 
          \biggr)\biggr) \;.
}                   
For the minimum $\mathcal{V}^* \simeq \tau_{\rm b}^{2/3}\simeq 92158$ of our model from section \ref{sec_gut_3} and $L$ of the order $L=40\ldots50$,
the prefactor in \eqref{quintessence} is of the right order of magnitude.

\begin{figure}[p]
\centering
\renewcommand{\subfigcapskip}{-0pt}
\subfigure[$V_F(\mathcal{V},\tau_1)$]{
\includegraphics[width=0.3\textwidth]{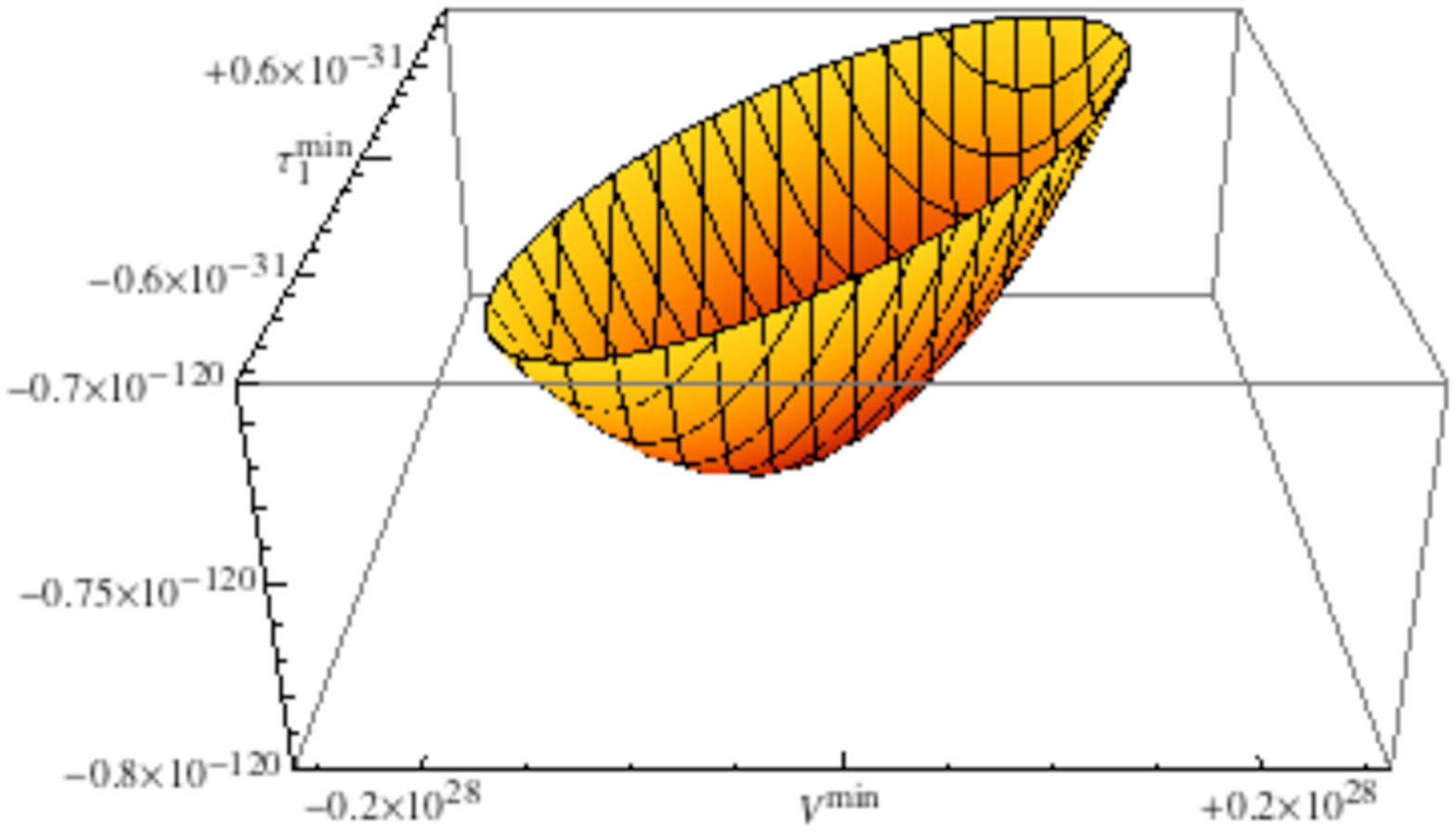}
}
\hfill
\subfigure[$V_F(\mathcal{V},\tau_2)$]{
\includegraphics[width=0.3\textwidth]{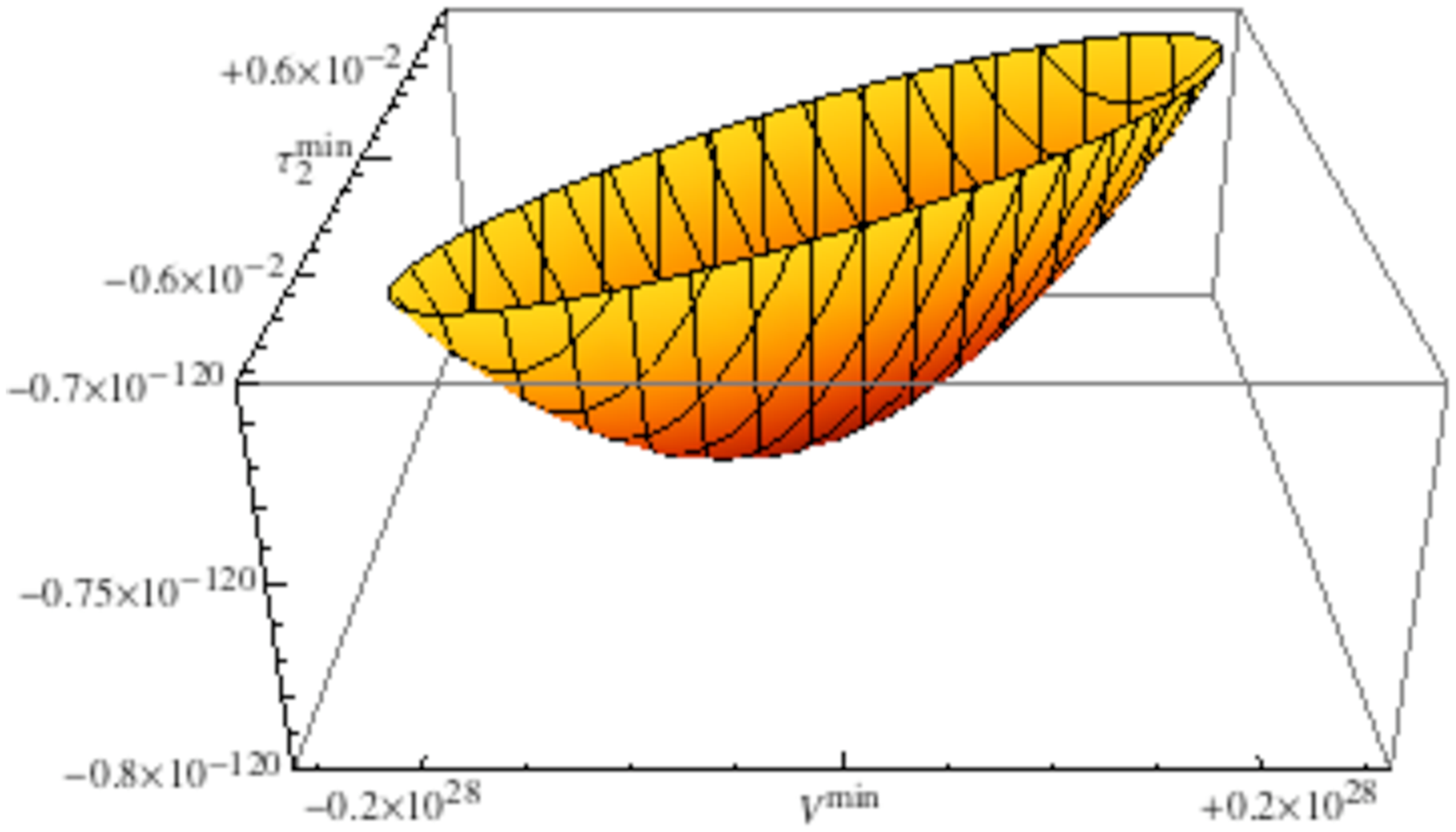} 
}
\hfill
\subfigure[$V_F(\tau_1,\tau_2)$]{
\includegraphics[width=0.3\textwidth]{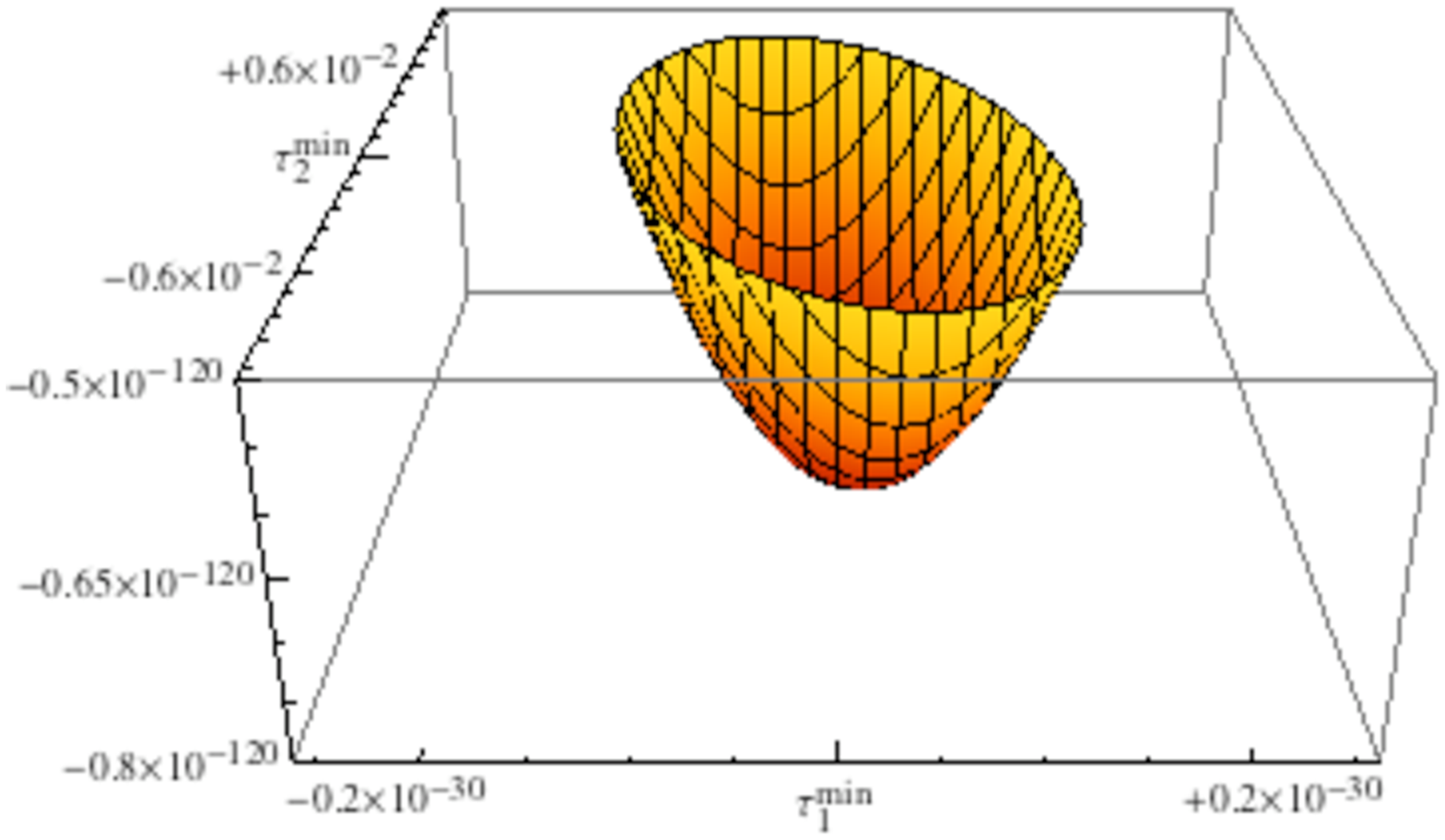} 
}

\vspace{-5pt}
\caption{\small F-term potential of the 
$\Lambda$ model 
in the vicinity of the minimum.\label{figcc}}
\vspace{-11pt}
\end{figure}

\bigskip
Although the GUT models from the previous section may contain a quint\-essence field, let us now take a different point of view.
Since in our scenario we have exponential control over 
$W_0^{\rm eff}$ and $A^{\rm eff}$,  one might 
ask whether it is possible to dynamically
find a minimum 
$ V_F^* \simeq - \lvert W^{\rm eff}_0\rvert^2\, \mathcal{V}^{-3}  
\simeq - 10^{-120}M_{\rm Pl}^4$ realised without fine-tuning.
Ignoring the weak scale for a moment, for the string scale
we have $M_{\rm s}\sim\mathcal{V}^{-1/2}M_{\rm Pl}>{\rm TeV}$ which
implies that $\mathcal{V}<10^{30}$. To keep the tuning of 
$a,b,\mathcal{A},\mathcal{B}$ moderate, we identify $\lvert W_0^{\rm eff}\rvert\sim10^{-15}$ and thus $\mathcal{V}\sim 10^{30}$ 
as a natural choice to 
realise $ V_F^*  
\simeq - 10^{-120}M_{\rm Pl}^4$. A set of parameters dynamically leading to such values
is for instance
\begin{equation*}
\arraycolsep1.5pt
\begin{array}{@{}lcllcllcllcllcllcl@{}}
  \mathcal{A}  &=& 1\,,            \;\;&
  \mathcal{B}  &=& 0.1\,,            \;\;&    
  \mathcal{C}_1&=& 1\,,              \;\;&
  \mathcal{C}_2&=& 3\,,              \;\;&   
  a            &=& \frac{2\pi}{13}\,, \;\;&
  b            &=& \frac{2\pi}{14}\,,     \\[4pt]
  g_s          &=& \frac{1}{5}\,,    \;\;&   
  \eta_{\rm b} &=&  1\,,             \;\;&
  \eta_1       &=& \frac{1}{30}\,,   \;\;&
  \eta_2       &=& \frac{1}{6}\,,    \;\;&
  \chi         &=& \multicolumn{4}{l}{-452\,, }    
\end{array}
\end{equation*}
and the plots showing the potential in the vicinity of the minimum can be found in figures \ref{figcc} on page \pageref{figcc}. The numerical
values specifying the minimum are
\eq{
  \mathcal{V}^* = 6.4\cdot10^{28}\;,\quad
  T_1^* = 68.84\;,\quad
  T_2^* = 11.53\;,\quad
  V_F^*= -7.8\cdot 10^{-121}M_{\rm Pl}^4\;,
}
and because the AdS minimum is at $V_F^*\sim-10^{-120}M_{\rm Pl}^4$,
the warp factor $a=10^{-15}$ in the uplift potential \eqref{uplift}
does not involve any fine-tuning of the flux parameters $K$ and $M$.

We conclude that there exist  vacua of the scalar potential
\eqref{effspot} whose tree-level cosmological
constant has the right order of magnitude.
However, this clearly does not solve the cosmological constant
problem, as we have not yet identified the Standard Model and the origin of
the weak scale.
Once we try to introduce the MSSM into this set-up, we are confronted with
the usual problems. Let us briefly explain three possibilities:
\begin{itemize}
\setlength{\itemsep}{0pt}

\item Localising the MSSM on $D7$-branes wrapping the cycles $\Gamma_{(1,2)}$
leads to soft masses below  the gravitino mass scale
$M_{3/2}\simeq 10^{-18}\,{\rm eV}$ which itself is ridiculously small.

\item Since $M_{\rm s}\sim {\rm TeV}$,
we could break supersymmetry at the string scale
and place a non-supersymmetric anti $D3$-brane configuration
realising the \linebreak[4] MSSM  at the bottom
of a throat, i.e. on the TeV brane in the RS scenario. 
The uncancelled NS-NS tadpole of the non-supersymmetric brane
configuration would be the  
red-shifted uplift term. However, all mass scales 
in the throat are red-shifted as well
\cite{Randall:1999ee} so that the stringy excitations such as
squarks have masses 
$M_{\tilde Q}\simeq a\, M_{\rm s}\simeq \Lambda^{1/4} = 10^{-3}\, {\rm eV}$.

\item A third option is to place an explicitly supersymmetry
breaking D-brane configuration in the bulk, i.e. 
on the Planck brane in the RS scenario. Then the
superpartners have string scale masses in the TeV region, 
but we get an additional positive contribution
$V\sim \mathcal{O}(M_{\rm s}^4)$
to the scalar potential.

\end{itemize}

In conclusion, even though we have exponential control over the effective 
parameters $W_0^{\rm eff}$ and $A^{\rm eff}$,
the cosmological constant problem  is not even touched. 
It can be phrased 
as the problem of hiding the TeV scale supersymmetry breaking of
the Standard Model such that it
does not induce a large contribution to the tree-level value $\Lambda_0$.


\section{Summary and Conclusions}
\label{sec_concl}

In this paper, we have studied string theory compactifications of type IIB 
orientifolds where the complex structure moduli are assumend to be stabilised by fluxes such 
that $W_{\rm flux}\rvert_{\rm min.}=0$. In addition, we considered a compactification manifold of 
{\em swiss-cheese} type where gaugino condensation on two stacks of 
$D7$-branes leads to a race-track superpotential. Taking into account 
 instanton corrections to the gauge kinetic function manifesting
themselves as
poly-instanton corrections to the race-track superpotential,
we constructed a 
scenario featuring exponential control over the parameters $W_0^{\rm eff}$ 
and $A^{\rm eff}$ in an effective superpotential of the form 
$W^{\rm eff}=W_0^{\rm eff}-A^{\rm eff} \exp(-a\, T)$. Such superpotentials are used 
for moduli stabilisation in KKLT and LARGE Volume Scenarios. However, in our setup we can arrange for exponential small values of these parameters without 
fine-tuning.

\bigskip
Using this setup, we were able to find minima of the resulting scalar potential
realising  supersymmetric GUT scenarios 
featuring $M_{\rm s}=M_X\simeq 1.2\cdot 10^{16}\,{\rm GeV}$, $\alpha^{-1}\simeq\alpha^{-1}_X\simeq25$ 
and a gravitino mass in the TeV region without fine-tuning. However, despite 
the phenomenological interesting value of $M_{3/2}$, the soft terms in our 
first setup are strongly suppressed. The reason is that the race-track 
modulus is stabilised in a nearly supersymmetric minimum giving F-terms which are much smaller than expected. 
We proposed two resolutions of this issue and constructed the corresponding 
setups:
\begin{itemize}

\item First, we scaled our parameters such that effectively $W_0^{\rm eff}$ is 
scaled by a factor of $10^6$ leading to a larger gravitino mass but also to 
larger soft masses. The gaugino masses are dominantly generated by 
anomaly-mediation while the scalar masses are generated by gravity mediation 
leading to a stringy realisation of split supersymmetry
if the Higgs and Higgsino masses are tuned to small values.

\item The second possibility we considered was to generate the soft terms 
not by the F-term of the race-track modulus but also by the
F-term of the  small LVS K\"ahler modulus. 
The gaugino as well as the scalar masses are then generated by gravity 
mediation similarly to the original LARGE Volume Scenario.
The lightest modulus was on the edge of posing problems
with cosmology (CMP) and the $\mu$-problem could be solved
by the Giudice-Masiero mechanism.

\end{itemize}
Clearly, a more detailed analysis of the phenomenological
implications of these setups along the lines
of \cite{Conlon:2007xv,Aparicio:2008wh} would be very interesting.
Moreover, it remains to be seen whether one can indeed
construct global string or F-theory compactifications
realising all the features we assumed for this scenario.
Not the least challenge  is to concretely evade
the problem of intrinsic tension between moduli stabilisation
via instantons and a chiral MSSM sector raised in
\cite{Blumenhagen:2007sm}.

\bigskip
Finally, in the last section we mentioned that for our GUT models there is, similarly to the LARGE Volume Scenarios, a natural candidate for a quintessence field. Taking a different point of view, we were also able to
construct a setup with tree-level 
cosmological constant at the order of $\Lambda_0\sim -10^{-120} M^4_{\rm Pl}$ 
which can be uplifted to a positive value without fine-tuning. However, 
although we obtained an encouraging value for $\Lambda_0$, introducing the 
Standard Model in this setup will spoil this feature.


\subsection*{Acknowledgements}

We gratefully acknowledge discussions with
N. Aker\-blom, E. Dudas, T. Grimm, D. L\"ust, M. Schmidt-Sommerfeld
and T. Weigand. We also thank S. P. de Alwis for pointing out a possible unclarity in the first version of this paper.


\clearpage

\clearpage
\nocite{*}
\bibliography{references}
\bibliographystyle{utphys}


\end{document}